\documentclass{aastex62}

\usepackage[utf8]{inputenc}
\usepackage{graphicx}
\usepackage{amssymb}
\usepackage{subfigure}
\usepackage{numprint}
\usepackage{hyperref}
\usepackage{todonotes}
\usepackage{xcolor}
\usepackage{amsmath}
\usepackage{bm}

\DeclareMathOperator{\diag}{\mathrm{diag}}

\graphicspath{{./}{figures/}}

\received{November 2, 2018}
\revised{December 17, 2018}
\accepted{December 31, 2018}

\shorttitle{Bistatic full-wave radar tomography detects deep interior voids, cracks and boulders}
\shortauthors{Sorsa et al.}

\begin{document}

\title{Bistatic full-wave radar tomography detects deep interior voids, cracks and boulders in a rubble-pile asteroid model}

\correspondingauthor{Liisa-Ida Sorsa}
\email{liisa-ida.sorsa@tuni.fi}

\author[0000-0001-5663-0958]{Liisa-Ida Sorsa}
\affil{Laboratory of Mathematics,
Tampere University, PO Box 553, 33101 Tampere, Finland}

\author[0000-0002-7518-2473]{Mika Takala}
\affil{Laboratory of Mathematics,
Tampere University, PO Box 553, 33101 Tampere, Finland}

\author[0000-0002-5398-3640]{Patrick Bambach}\affil{Max Planck Institute for Solar System Research,  Justus-von-Liebig-Weg 3, 37077 Göttingen, Germany}

\author[0000-0001-8341-007X]{Jakob Deller}\affil{Max Planck Institute for Solar System Research,  Justus-von-Liebig-Weg 3, 37077 Göttingen, Germany}

\author[0000-0002-6184-7681]{Esa Vilenius}\affil{Max Planck Institute for Solar System Research,  Justus-von-Liebig-Weg 3, 37077 Göttingen, Germany}

\author[0000-0002-9131-9070]{Sampsa Pursiainen}
\affil{Laboratory of Mathematics,
Tampere University, PO Box 553, 33101 Tampere, Finland}

\begin{abstract}

In this paper, we investigate full-wave computed radar tomography (CRT) using a rubble-pile asteroid model in which a realistic shape (Itokawa) is coupled with a synthetic  material composition and structure model. The aim is to show that sparse bistatic radar measurements can distinguish details inside a complex-structured rubble-pile asteroid. The results obtained suggest that distinct local permittivity distribution changes such as surface layers, voids, low-permittivity anomalies, high-permittivity boulders, and cracks can be detected with bistatic CRT, when the total noise level in the data is around -10 dB with respect to the signal amplitude. Moreover, the bistatic measurement set-up improves the robustness of the inversion compared to the monostatic case. Reconstructing the smooth Gaussian background distribution was found to be difficult with the present approach, suggesting that complementary techniques, such as gravimetry, might be needed to improve the reliability of the inference in practice. 

\end{abstract}


\section{Introduction}\label{intro}

This paper investigates spaceborne computed radar tomography (CRT) applied to a rubble-pile asteroid model. Spaceborne CRT inherits from the airborne ground penetrating radar (GPR) which was originally developed to serve in cost-effective surveying of the underground in applications which entail the necessity to work with an antenna not in contact with the surveyed structure \citep{Catapano2012a, Catapano2012b,Soldovieri2017}. Such airborne GPR systems have been validated by airborne measurement data and numerical experiments \citep{fu2014}. Space mission concepts to perform radio tomography of small asteroids have been proposed by, for example, \citet{aspaugh2001}, \citet{safaeinili}, \citet{snodgrass2018}, and \citet{discus}. 

The first attempt to reconstruct the deep interior of a small solar system body (SSSB) was the Comet Nucleus Sounding Experiment by Radio-wave Transmission (CONSERT), a part of European Space Agency's (ESA) Rosetta mission to explore the comet 67P/Churyumov-Gerasimenko, in which a radio signal was transmitted between the orbiter \emph{Rosetta} and the lander \emph{Philae} \citep{kofman1998, Kofman2007, kofman2015}. Many more missions aiming at exploring the structure and composition of SSSBs are currently ongoing or being planned. In August 2018, the Osiris-REx mission by NASA \citep{osirisrex} begun its rendezvous with the asteroid 101955 Bennu to measure its physical, geological and chemical properties and collect a sample of the asteroid surface regolith \citep{Lauretta2017}. The Hayabusa mission by the Japanese Aerospace Exploration Agency (JAXA) explored the asteroid Itokawa already in 2005 \citep{hayabusa, okada, nakamura, tsuchiyama} and retrieved surface regolith for analysis on Earth \citep{fujiwara}, confirming the  Itokawa's Earth-based classification as an S-type asteroid. The images taken by Hayabusa on Itokawa have been used to analyze the size distribution of boulders on its surface. The results indicate that the boulders cannot solely be the  product of cratering but that many of them originated from the disruption of a larger body \citep{Saito2006, Michikami2008}. JAXA's mission Hayabusa2 arrived at the asteroid 16173 Ryugu in mid-June 2018. It  will survey its target for a year and a half, returning back to Earth in December 2020 \citep{tsuda}. These missions by NASA and JAXA concentrate on the surface properties of the target asteroids and do not carry instruments which could be used to explore the deep interior structures. 

The next candidate to deploy a CRT system to explore the deep interior of a SSSB is ESA's asteroid mission Hera which targets the binary near-Earth asteroid system Didymos and is due to launch in 2023 \citep{Michel2016}.  The current plan allows deployment of a 6U  CubeSat form factor payload \citep{carnelli2018} which has been recently proposed for the radar-carrying spacecraft concept DISCUS (Deep Interior Scanning CUbeSat) \citep{discus}. Furthermore, ESA's recent Concurrent Design Facility (CDF) study report \citep{cdfreport} suggests that small planetary platforms involving CubeSats can provide future opportunities for CRT.    

In this paper, CRT is  applied to a rubble-pile asteroid model  utilizing the mathematical approach proposed in   \citet{pursiainen2016} and \citet{farfield}. We investigate a synthetic framework featuring the radar specifications of the DISCUS concept, and a target with a complex shape, structure and material composition. The shape of the asteroid Itokawa is coupled with a Gaussian random field model for the relative dielectric permittivity distribution together with added structural details. We base our model on the recent observations and impact simulations which suggest that the internal porosity of the asteroid body varies, increasing towards the center, and that it may have a detailed structure  \citep{carry2012density,deller_hyper-velocity_2017,Jutzi2017}.  

The aim is to validate numerically the mathematical approach of \citet{pursiainen2016} and \citet{farfield} and also to explore, whether sparse bistatic radar measurements can distinguish details inside a complex-structured rubble pile asteroid. 

\section{Materials and Methods}
\label{methods}

In this study, we utilize the finite element time-domain (FETD) approach  \citep{pursiainen2016} equipped with a far-field model proposed in \citet{farfield}. In FETD, the radar signal is propagated over a suitably chosen time interval  within a volumetric finite element (FE) mesh which can be adapted accurately to a given set of surface features. The permittivity structure can be found via a multigrid approach \citep{multigrid}. That is, the signal is inverted using a coarser grid which is a nested strucure  with respect to the more refined FE mesh. 

\subsection{Model of the asteroid Itokawa} \label{sec:femmodel}

The FE mesh was constructed based on the detailed shape model of the asteroid Itokawa (Figure \ref{fig:itokawa_model}, left) which is openly available as a triangular stereolitography (STL) mesh \citep{itokawastlfile}. The scaling of the asteroid was assumed to match with the actual size of Itokawa, whose  longest diameter is 535 meters \citep{fujiwara}. The unstructured triangulated asteroid surface was imported to Meshlab \citep{meshlab} and resampled with a Poisson-disk sampling algorithm to obtain a uniform mesh with a desired resolution of  \numprint{5762} nodes and \numprint{11520} faces. The volumetric asteroid model was divided into a  surface and interior compartment. The boundary between these two parts was obtained by scaling a resampled and smoothed-out version of the outer surface by the factor of $0.9$. An additional ellipsoidal compartment was utilized as a void detail, when simulating the measurement data.

\subsection{The structure and the 3D domain structure}

\begin{figure}[!ht]
  \begin{center}   
   \includegraphics[width=0.22\textwidth]{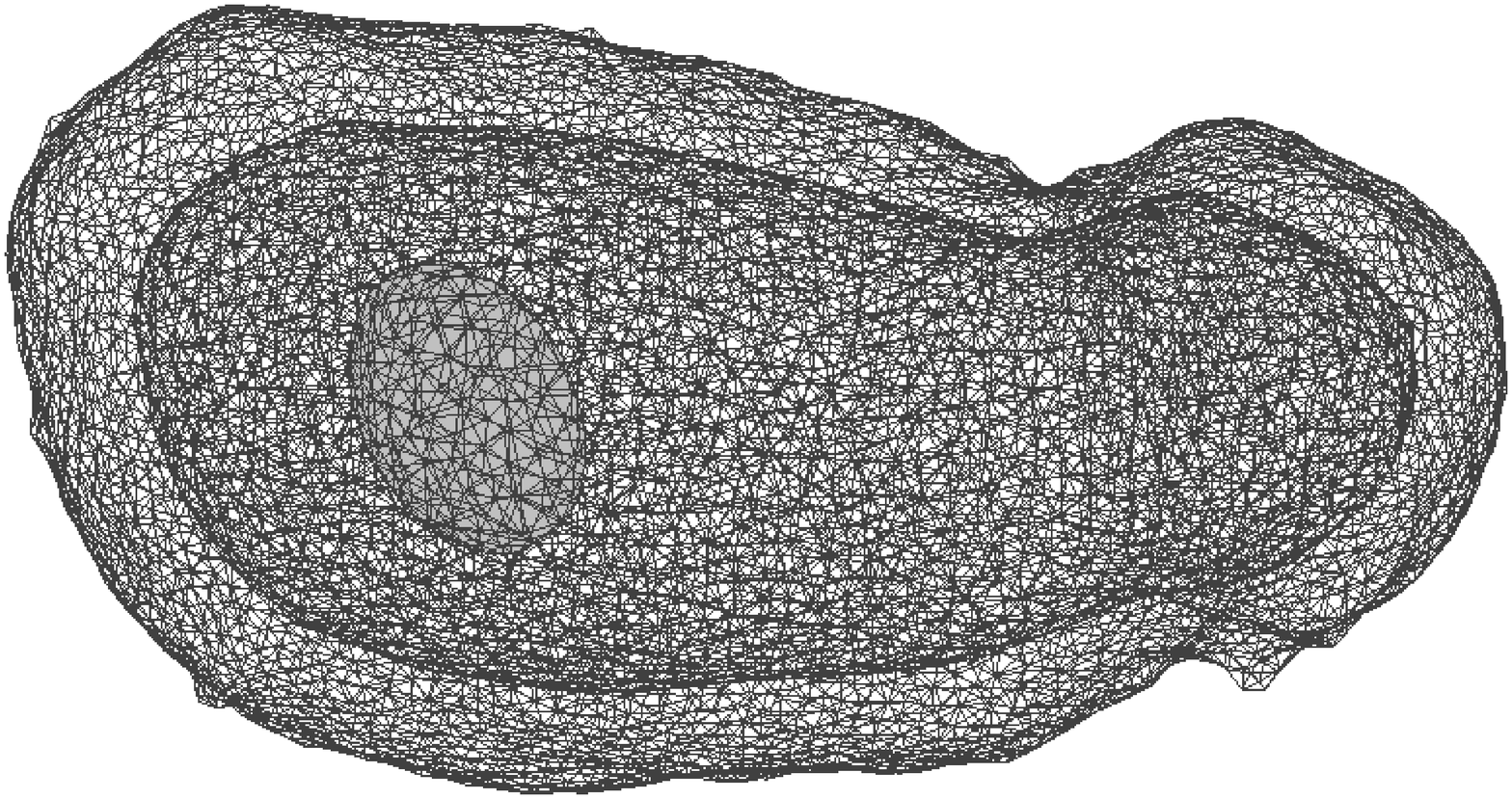} \hskip0.5cm
   \includegraphics[width=0.18\textwidth]{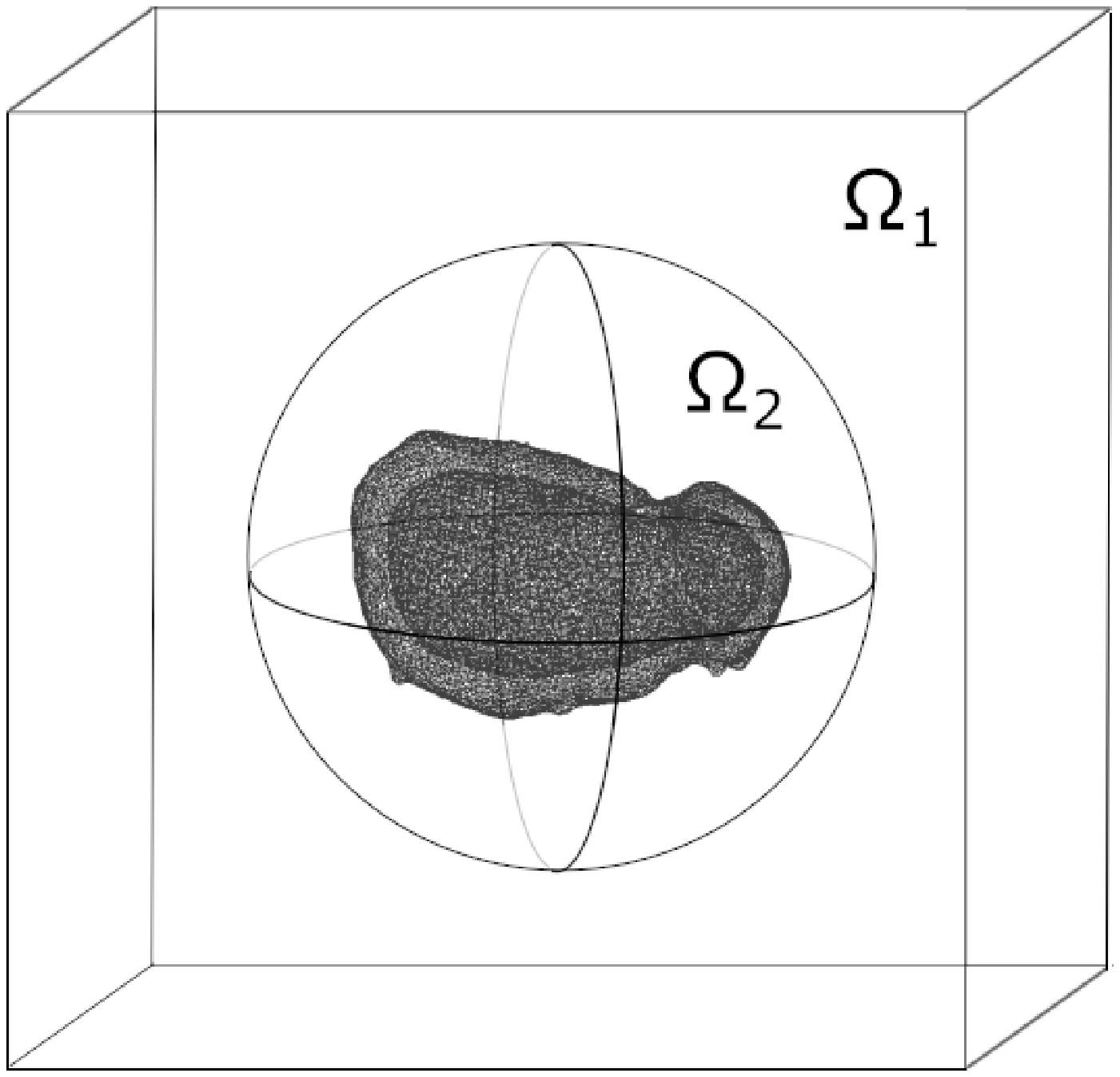}   \hskip0.5cm  \includegraphics[width=0.16\textwidth]{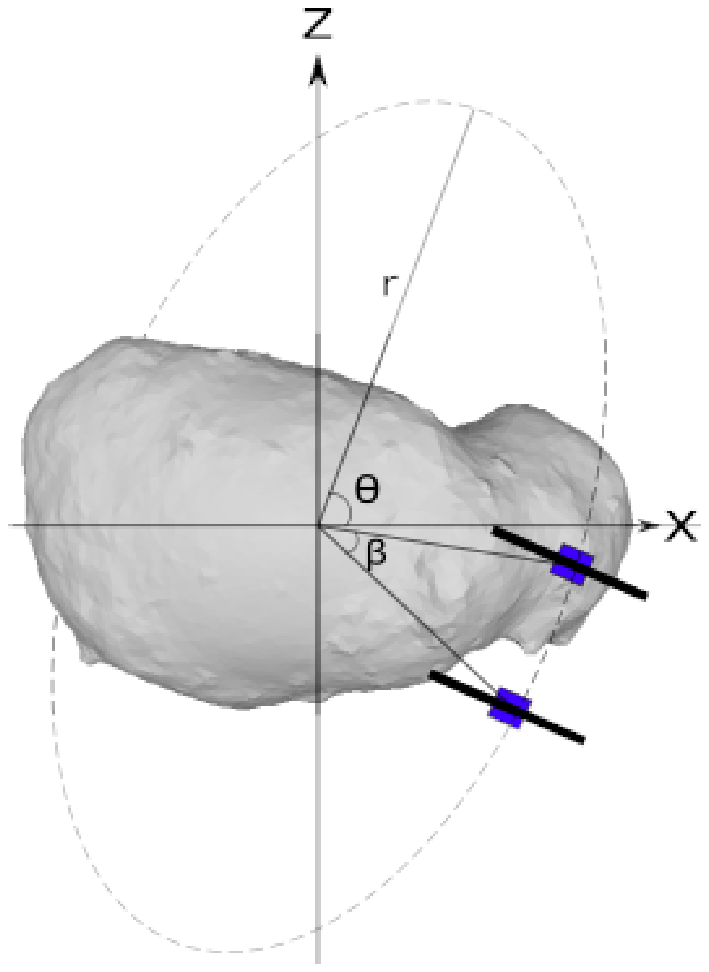} \hskip0.5cm
   \includegraphics[width=0.24\textwidth]{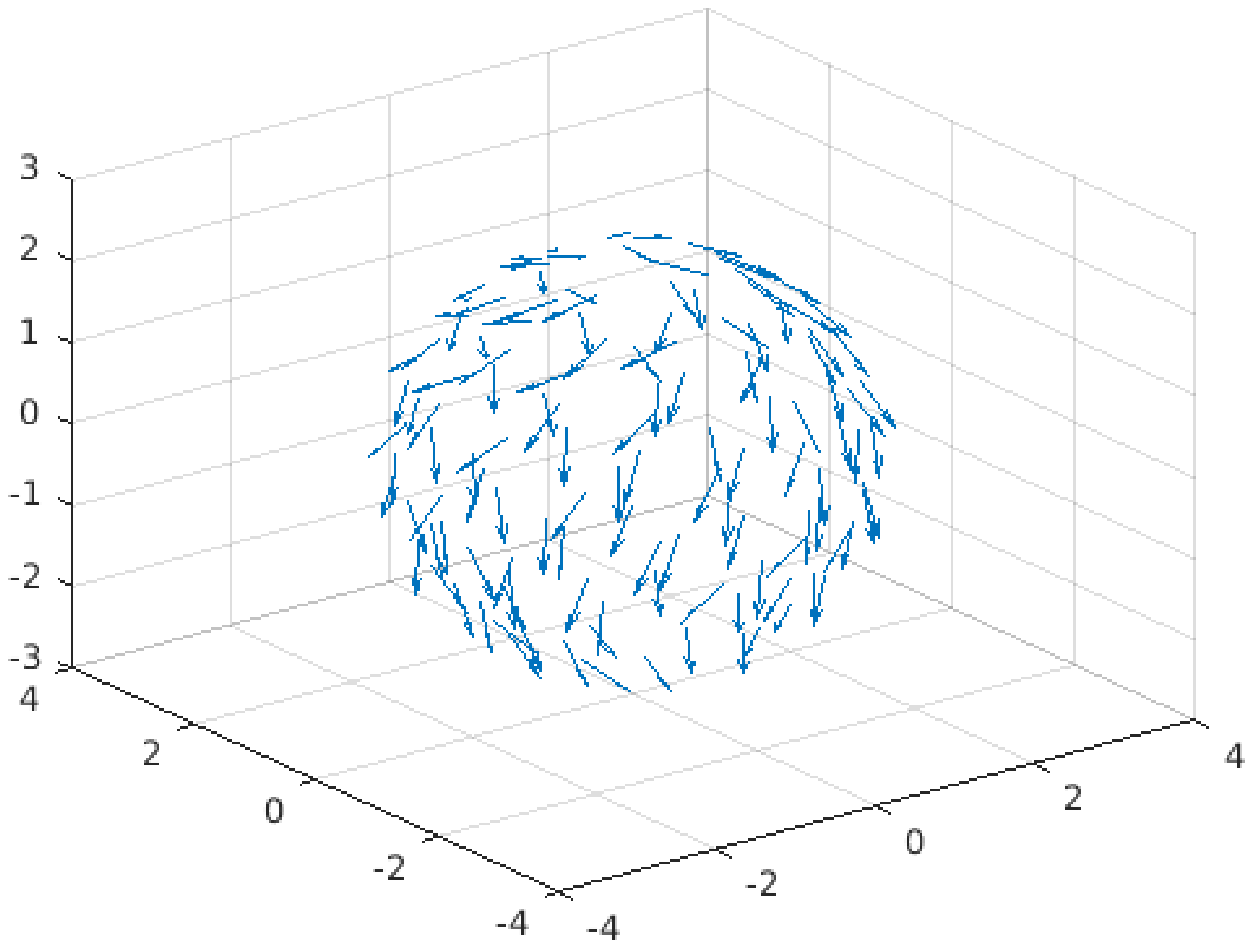} 
  \end{center}
  \caption{{\bf Left:} The FE model of Itokawa including a single deep interior void in the body of the asteroid. The detailed model includes surface and interior compartments and  an ellipsoidal  deep interior detail. \label{fig:itokawa_model}{\bf Second from the left:} The  computation domain $\Omega$ of this study comprises two nested subdomains $\Omega_{1}$ and  $\Omega_{2}$. The outer subdomain,  $\Omega_{1}$, contains a split-field perfectly matched layer to simulate open field scattering. The asteroid model is contained in $\Omega_{2}$. The spherical boundary between the subdomains was applied in simulating the far-field signal transmission and measurements as described in \citet{farfield}.  {\bf Third from the left:} A schematic illustration of the bistatic measurement approach. The angle $\theta$ depicts the limiting angle between the satellite orbiting plane normal and the asteroid spin. The angle $\beta$ is the angle between the orbiters in their orbitting plane. The two satellites orbit at distance $r$ from the center of the asteroid. {\bf Right:} A quiver plot of the 64 orbiter points and orientations around the asteroid.} \label{fig:satellitepoints}\label{fig:sparsepoints}\label{fig:malli}
\end{figure}

The computation domain $\Omega$, depicted in Figure \ref{fig:malli} (second from the left) was constructed in the Gmsh software \citep{gmsh}, in which the structure of the computation domain $\Omega$ can be described within a geometry (GEO) file and the compartments of the asteroid model can be obtained via their STL surface grids. Two coarse FE meshes with  \numprint{38}k nodes and \numprint{212}k tetrahedra were generated using Gmsh. One of these was employed for simulating the measurements, and the other one, not including any interior details, was used in the inversion stage. The reason to  apply two different grids was to avoid the so-called {\em inverse crime}, i.e., overly good match between the actual permittivity distribution and its reconstruction. The permittivity was modeled as a piecewise constant function within the coarse grid. A nested two times uniformly refined grid was utilized, when simulating the signal propagation, in order to achieve a sufficient resolution with respect to the wavelength.

\subsection{The set of measurement points}

The measurement point set was modeled after the DISCUS mission concept \citep{farfield} in which a master and slave CubeSat both  equipped with a half-wavelength dipole antenna orbit the target asteroid at a 5 km distance. The angle between the orbiters is 25 degrees with respect to the asteroid's center of mass (Figure \ref{fig:satellitepoints}, third from the left). The measurement configuration is bistatic: the master CubeSat both transmits and records the signal, and the slave  provides an additional receiver. The angle $\theta$ between the measurement plane normal and the asteroid spin, determining the angular coverage of the measurements, is assumed to be $70$ degrees. Optimally, a full coverage could be obtained with $\theta  = 90^{\circ}$. A practically obtainable orbiting direction may be expected to have a somewhat but not very much  lower value, as the target asteroid will be likely to have a close-to retrograde spin \citep{laspina2004}. In total, 64 measurement points were included in the point cloud. The resulting bistatic set of measurement points and antenna orientations has a sparse and limited-angle spatial coverage  with an aperture around the z-axis (Figure \ref{fig:sparsepoints}, right).

\subsection{Radar specifications}

Inheriting from the DISCUS concept \citep{discus}, the radar is assumed to have a transmission power of 10 W,  a \mbox{2 MHz} total signal bandwidth, and a relatively low 20 MHz   center frequency. In practice,  $\leq 100$ MHz  will be necessary in order to achieve appropriate signal penetration and to minimize solar noise \citep{kofman2012}. The maximal range (imaging) resolution following from these parameters is about 35-40 m inside an asteroid with relative permittivity approximately 4. 

\subsection{Computing Gaussian random field inside a given asteroid geometry}

\begin{figure}[!ht] 
  \centering 
\includegraphics[width=0.4\textwidth]{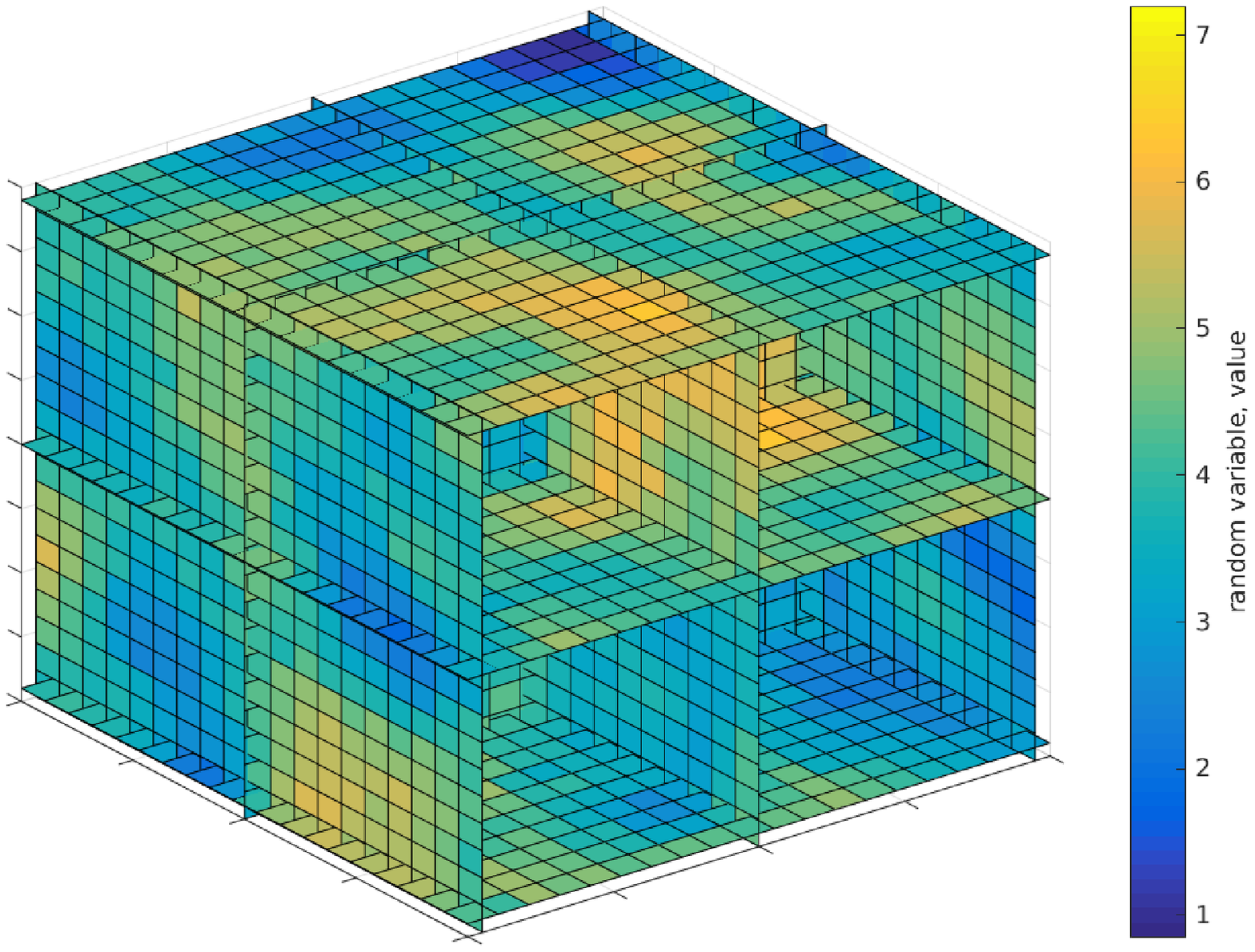} \hskip0.5cm
  \includegraphics[width=0.48\textwidth]{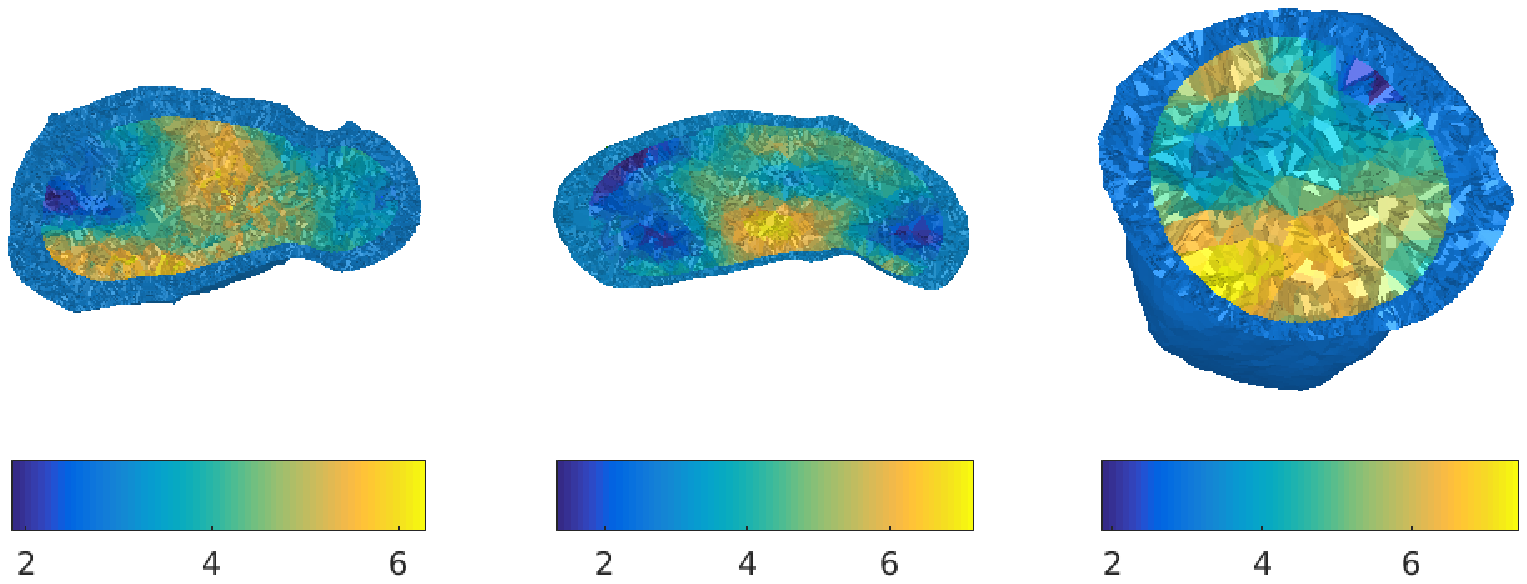}
\caption{{\bf Left:} A Gaussian random field within a regular  20-by-20-by-20 grid. {\bf Right:} In each model (A)--(E), the smooth interior part corresponds to a Gaussian random field. The surface layer is associated with a constant and comparably lower permittivity value $\varepsilon_r = 3$. The figures from left to right, show a slice cut-view of the permittivity distribution in the $xy$-, $yz$- and $zx$-plane, respectively.}\label{fig:asteroid_interior}
\label{fig:randomfieldgrids}
\end{figure}

A Gaussian random field for the dielectric relative permittivity, $\varepsilon_r$, was first generated in a regular 20-by-20-by-20 lattice (Figure \ref{fig:randomfieldgrids}, left). The mean of the random field was set to $\varepsilon_r = 4$, assuming a  standard deviation equal to one. The correlation value between the adjacent lattice points, was set to $0.2$ based on visual examination. The correlation was chosen to be isotropic.  These choices were thought to roughly account for the recent impact simulation results in which large pieces of porous rubble, bound together by gravity, are concentrated in the interior \citep{Jutzi2017}, yet allowing some clear randomness of the distribution due to the fact that the exact interior structure is mainly unknown  \citep{carry2012density}. 

Implementation of the Gaussian random field algorithm was retrieved from \citet{randomfieldmfile}. The asteroid interior geometry was then fitted inside the cubic mesh and the values were assigned to the finite element mesh nodes via the nearest neighbour interpolation principle. Each finite element mesh tetrahedron was assigned a value which is the average of the electrical permittivities of each node constituting the element. Finally, the random field was restrictred to  permittivity values greater or equal to one, which is the dielectric permittivity of vacuum and, therefore, the physical minimum of relative permittivity. 

\subsection{Structural details}

This section describes the details of the five asteroid models (A)--(E) applied in this study. The details of these have been given also in Table \ref{model_table}. In each model, the permittivity distribution in the interior part is determined by a Gaussian random field with values approximately between two and six (Figure \ref{fig:asteroid_interior}, right). Furthermore, each one includes a  surface layer with a lower permittivity to model the higher porosity near the surface predicted in the impact simulations \citep{Jutzi2017}. In each case, a different random realization of the Gaussian random field is used.

\begin{deluxetable*}{llllccc}[b!]
\tablecaption{Details in the asteroid models (A)--(E) \label{model_table}}
\tablecolumns{7}
\tablenum{1}
\tablewidth{0pt}
\tablehead{
\colhead{} & \colhead{} & \colhead{} & \colhead{} & 
\multicolumn{3}{c}{Detail size\tablenotemark{a}}\\
\colhead{Model} &
\colhead{Description} &
\colhead{Detail} & \colhead{$\varepsilon_r$} & \colhead{Length } & \colhead{Width} & \colhead{Depth} 
}
\startdata
(A) & Single  void & Ellipsoid & 1 & 120 m & 120 m & 55 m\\
& & & & 8.0$\lambda$ & 8.0$\lambda$ & 3.7$\lambda$ \\
& & & & 0.8$\lambda$ & 0.8$\lambda$ & 0.4$\lambda$ \\
(B) &  Highly porous inclusion & Ellipsoid & 2  & 120 m & 120 m & 55 m\\ 
& & & &  11.3$\lambda$ & 11.3$\lambda$ & 5.2$\lambda$ \\
& & & &  1.1$\lambda$ & 1.1$\lambda$ & 0.5$\lambda$ \\
(C) &  High-permittivity boulder & Ellipsoid & 15 & 120 m & 120 m & 55 m\\ 
& & & &  31.8$\lambda$  & 31.8$\lambda$  & 14.1$\lambda$  \\
& & & &  3.1$\lambda$  & 3.1$\lambda$  & 1.4$\lambda$  \\
(D) & Deep crack & Crack & 1 & 185 m & 10--55 m & 10--50 m\\
& & & &  12.3$\lambda$  & 0.7--3.7$\lambda$  & 0.7--3.3$\lambda$  \\
& & & &  1.2$\lambda$  & 0.07--0.4$\lambda$  & 0.07--0.3$\lambda$  \\
(E) & Shallow crack & Crack & 1 & 185 m & 10--55 m & 10--50 m\\
& & & &  12.3$\lambda$  & 0.7--3.7$\lambda$  & 0.7--3.3$\lambda$  \\& & & &  1.2$\lambda$  & 0.07--0.4$\lambda$  & 0.07--0.3$\lambda$  \\
\enddata
\tablenotetext{a}{Sizes are indicated in meters, with respect to the wavelengths of a 20 MHz center frequency signal, and with respect to the wavelengths of a 2 MHz bandwidth signal.}
\tablecomments{The respective center wavelengths of the 20 MHz signal are $\lambda = \{15.0, 10.6, 3.9\}$ meters in each of the relative permittivities $\varepsilon_r = \{1,2,15\}$. The wavelengths of the 2 MHz bandwidth signal in the respective $\varepsilon_r$ values are $\lambda=\{150, 106, 39\}$ meters.}
\end{deluxetable*}

\subsubsection{Model (A): A single void} 

\begin{figure}[!ht] 
\begin{center}

\subfigure[Model (A):  3D model of the vacuum void.] {\label{fig:model_a}
\includegraphics[width=0.3\textwidth]{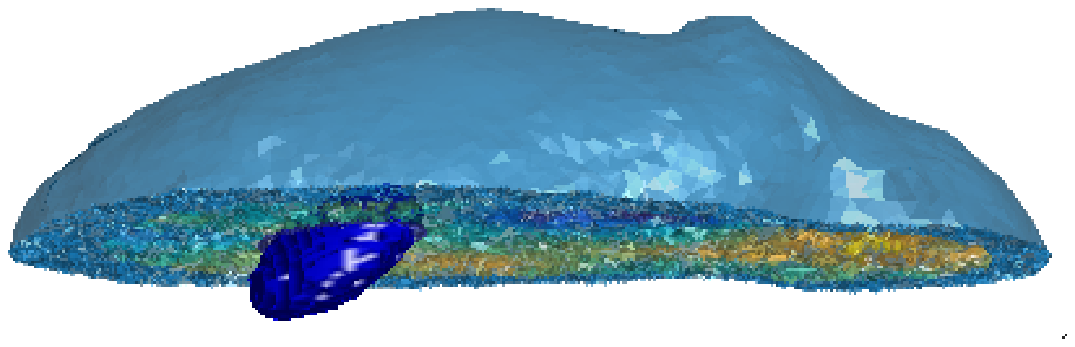}}
\subfigure[Model (A): Cut-view of the vacuum void model. The same structure with is used also in the models (B) and (C).] {\label{fig:void_slicefig}
\includegraphics[width=0.45\textwidth]{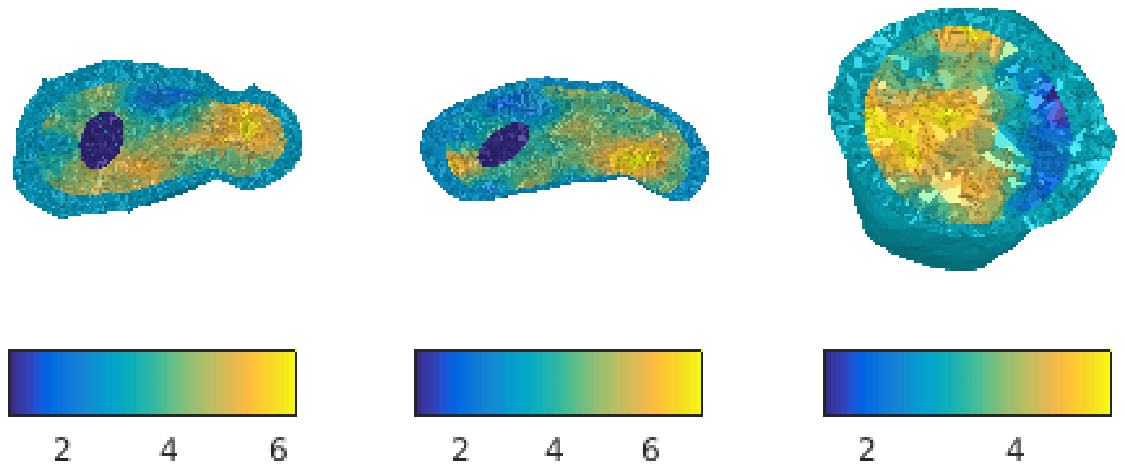}}\\
\subfigure[Model (B):  3D model of the highly porous inclusion.] {\label{fig:model_b} 
\includegraphics[width=0.3\textwidth]{exact_model_3d_void2.eps}} \hspace{2cm}
\subfigure[Model (C): 3D model of the high-permittivity boulder]{\label{fig:model_c}
\includegraphics[width=0.3\textwidth]{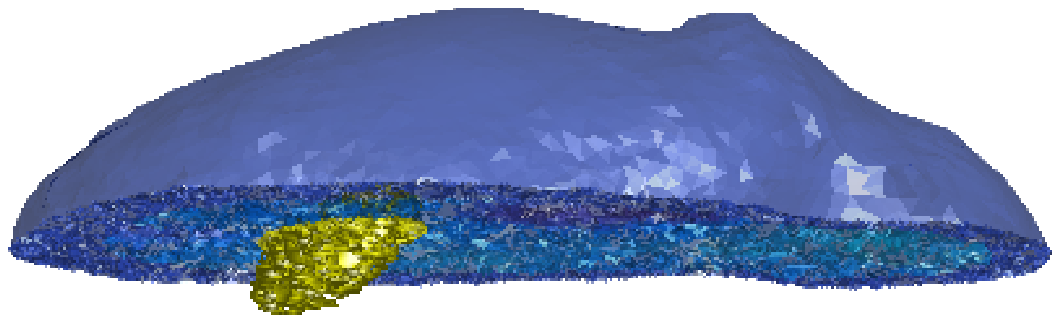}}

\subfigure[Model (D): Deep crack.]{\label{fig:model_d_3d}
\includegraphics[width=0.3\textwidth]{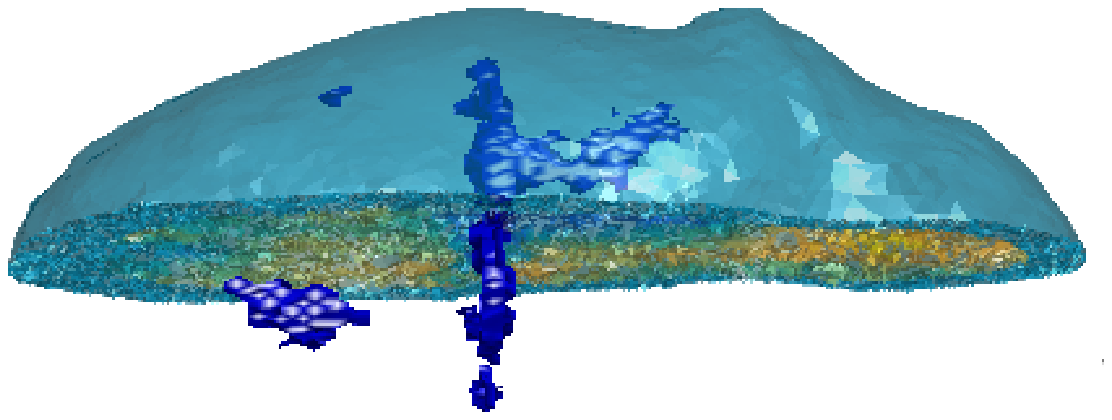}}
\subfigure[Model (D): A cut-view of the deep crack model.]{\label{fig:model_d}\includegraphics[width=0.45\textwidth]{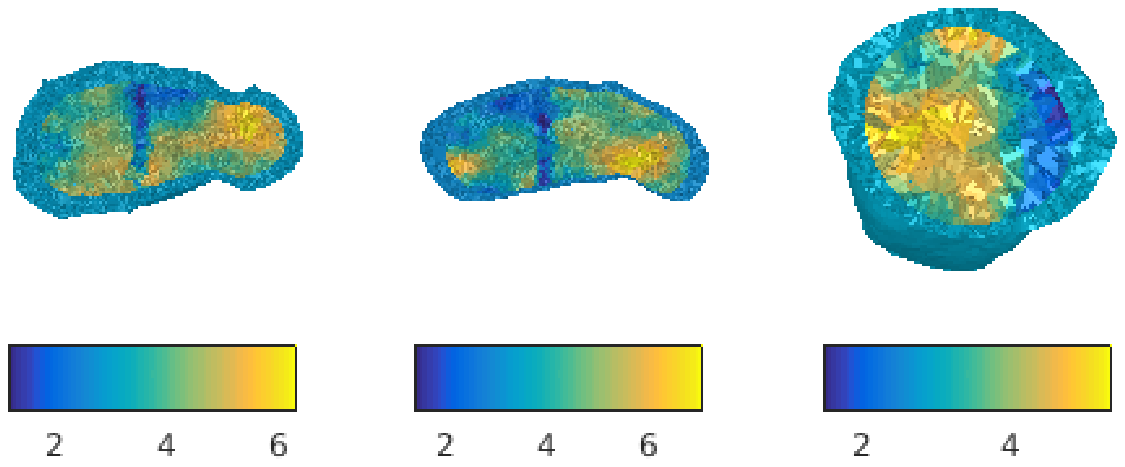}} 

\subfigure[Model (E): Shallow crack.]{\label{fig:model_e_3d} \includegraphics[width=0.3\textwidth]{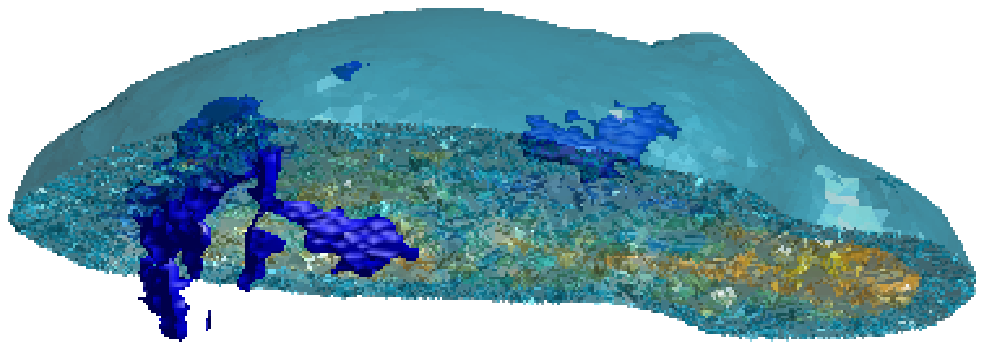}}
\subfigure[Model (E): A cut-view of the shallow crack model.]{\label{fig:model_e}\includegraphics[width=0.45\textwidth]{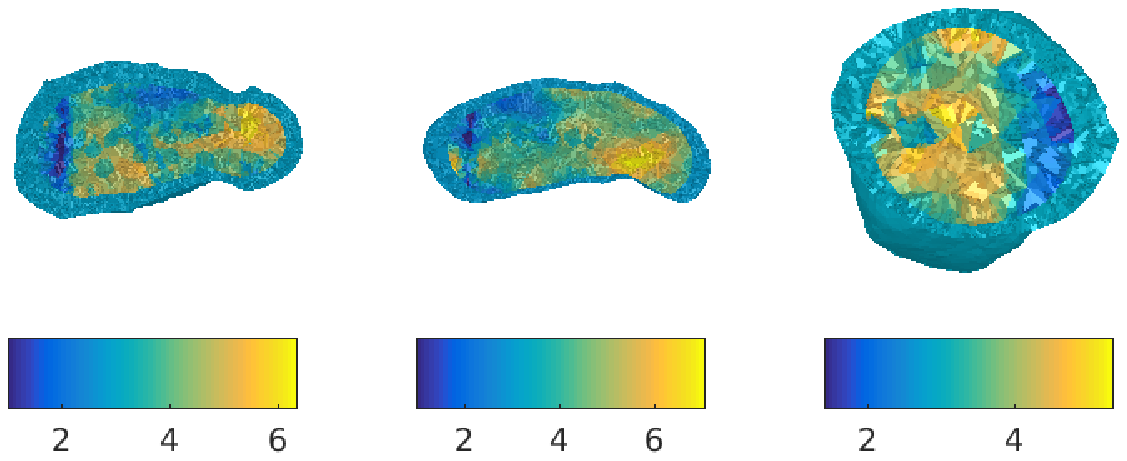}} 

\end{center}
\label{fig:shallowcrackfigs}
\caption{  {\bf Top row:} (a) Model (A) with deep interior vacuum void. The 3D model shows the ellipsoid shape of the anomaly. The cut-view (b) of the model on the right details the surface layer with $\varepsilon_r = 3$ and the Gaussian random field enclosing the ellipsoidal void with $\varepsilon_r = 1$. {\bf Second row:} 3D models of the models (B) and (C) in (c) and (d), respectively. The structures of these models are the same as the one in the top row (A), but the relative permittivities of the ellipsoids are different. {\bf Third row:} Illustration of the deep crack model (D) with the 3D model (e) and the cut-view illustration (f). {\bf Bottom row:} Illustration of the shallow crack model in 3D (g) and cut-view (h) showing the complicated shape of the crack.}
\label{fig:crackfigs}
\label{fig:void_slice}
\end{figure}

In (A), to investigate anomaly detection, a single ellipsoid with a relative permittivity value $\varepsilon_r = 1$ (vacuum) was placed deep in the interior (Figures \ref{fig:model_a} and  \ref{fig:void_slicefig}). This anomaly could be, for example, a void between two large boulders constituting a part of the body of the asteroid.    

\subsubsection{Model (B): A Highly porous inclusion}

In (B), we  study a situation in which the anomaly is not as distinct from the surroundings as a void space (Figure \ref{fig:model_b}). That is, the interior may include areas which are occupied by very porous or otherwise low permittivity material. Therefore, we investigate also an ellipsoid, whose relative dielectric permittivity is adjusted to $\varepsilon_r = 2$ instead  that of of the vacuum ($\varepsilon_r = 1$). 

\subsubsection{Model (C): A High-permittivity boulder}

An asteroid may also be formed around a single large boulder with dielectric permittivity significantly higher than that of its  surroundings. Such system is investigated with model (C) in which the dielectric permittivity of the enclosed ellipsoid is assigned with  the value $\varepsilon_r = 15$ which is significantly higher than the surrounding other material (Figure \ref{fig:model_c}). This value corresponds, e.g., to solid basalt \citep{hansen1973dielectric}, and was included in the study to investigate whether such permittivity structures can be reconstructed and to provide a more complete view of the capabilities of radar tomography.

\subsubsection{Model (D): A Deep crack}\label{sec:crack}

In (D), to model a random crack inside the asteroid, a simple nearest-neighbour walking algorithm was applied within the set of FE mesh nodes. Each node included in the path created by the random walk was assigned with the permittivity value of one (vacuum), and the permittivity of each tetrahedron in the FE mesh was determined by the average permittivity of its nodes. 
 
The resulting crack is irregularly shaped and smaller in comparison to the void (Figures \ref{fig:model_d_3d} and \ref{fig:model_d}). The width of the crack is in the order of 10-50 meters, depending on the location within the asteroid.

\subsubsection{Model (E): A Shallow crack} \label{sec:shallowcrack}

In (E), the random walk approach was utilized to create a shallow crack. This crack was also allowed to bifurcate to form a two-branched structure (Figures \ref{fig:model_e_3d} and  \ref{fig:model_e}).

\subsection{Conductivity distribution}

The conductivity distribution causing signal attenuation was assumed to be a latent parameter, i.e., an uninteresting nuisance parameter, depending on the permittivity according to $\sigma = 5 \xi \varepsilon_r$,  where $\xi = (\mu_0 / \varepsilon_0)^{-1/2}$ s$^{-1}$ with spatial scaling parameter $s = 2100$ m \citep{farfield}. If $\varepsilon_r = 4$, this results in a conductivity of around $2.5\times 10^{-5}$ S m$^{-1}$, which matches roughly with an attenuation rate of 25 dB km$^{-1}$  approximating a low-frequency signal decay within a porous body \citep{kofman2012}.

\subsection{Forward model}

The inversion process requires a reference permittivity $\varepsilon^{(bg)}_r$, a background model. Such a model is constructed similar to the detailed, structural models with the difference that the interior permittivity is constant, here $\varepsilon^{(bg)}_r = 4$. We utilize a linearized forward model given by \citet{pursiainen2016}:
\begin{equation}
    \mathbf{y} = \mathbf{Lx} + \mathbf{y}^{(bg)} + \mathbf{n}. 
\label{eq:forwardmodel}
\end{equation}
Here, the vectors $\mathbf{y}$ and $\mathbf{y}^{(bg)}$ contain the measured and simulated data for the actual permittivity distribution $\varepsilon_r$ and its constant-valued approximation $\varepsilon_r^{(bg)}$, respectively. The vector $\mathbf{x}$ determines the discretizised permittivity distribution, $\mathbf{L}$ denotes the Jacobian matrix resulting from the linearization at  $\varepsilon_r^{(bg)}$, and $\mathbf{n}$ contains both the measurement and forward modeling errors. 

The noise vector $\mathbf{n}$ is here assumed to be a zero-mean  Gaussian random variable due to the various potential and unknown noise formation processes, e.g., signal attenuation due to latent factors. Following from the rough estimates for the cosmic background noise level with the present radar and mission specifications \citep{farfield}, the standard deviation was chosen to be -15 dB with respect to the maximal entry of the difference $|\mathbf{y}-\mathbf{y}^{(bg)}|$ between the measured and simulated signal. 

\subsection{Inversion approach} 

We use a simple total variation (TV) based iterative regularization approach presented in \citet{pursiainen2016}. An estimate of $\mathbf{x}$ in equation \ref{eq:forwardmodel} can be produced via the iterative regularization procedure 

\begin{equation}
\mathbf{x}_\ell = (\mathbf{L}^T\mathbf{L} + \alpha \mathbf{D}\bm{\Gamma}_{\ell} \mathbf{D})^{-1}\mathbf{L}^T\mathbf{y}, \quad \bm{\Gamma}_{\ell} = \diag(|\mathbf{Dx}_{\ell}|)^{-1},
\label{eq:iterative}
\end{equation}

in which $\bm{\Gamma}_0 = \mathbf{I}$ and $\mathbf{D}$ is of the form:

\begin{align}
D_{i,j} = \beta\delta_{i,j}+\frac{\ell^{(i,j)}}{\max_{i,j}\ell^{(i,j)}}(2\delta_{i,j}-1), \nonumber \\
\delta_{i,j} = \begin{cases} 1, \quad \text{ if } j=i, \\ 0 \quad \text{ otherwise. } \end{cases}
\end{align}

The first term is a weighed identity operator limiting the total magnitude of $\mathbf{x}$, whereas the second term penalizes the jumps of $\mathbf{x}$ over the edges of the mesh, the $\ell^{(i,j)}$ being the edge length.

The inversion process in the equation \ref{eq:iterative} minimizes the function
$
F(\mathbf{x}) = \| \mathbf{Lx} - \mathbf{y}_{bg} -\mathbf{y} \|_2^2 + 2\sqrt{\alpha}\| \mathbf{Dx} \|_1, 
$
in which the second term equals the total variation of $\mathbf{x}$ if $\beta = 0$. The coefficients $\alpha$ and $\beta$ are the reqularization parameters.

The total variation penalty function is evaluated with respect to the tetrahedral inversion mesh as shown in \citet{farfield}. The inversion computation was run with regularization parameters values $\alpha = 0.01$ and $\beta =  \{0.005,0.0075\}$   depending on the model. The parameters were adjusted based on preliminary experiments with the goal to maximize the distinguishability of the interior details and to obtain an appropriate range of permittivity values with respect to the actual distribution. 

\section{Results}
\label{results}

\begin{figure}[!ht] 
\begin{center}
\subfigure[Anomaly deep in the interior. In addition to the background signal ${\bf y}^{(bg)}$ (red), the actual signal ${\bf y}$ is visualized for the model (A) = Void, (B) = Cavity and (C) = Boulder (blue, magenta and cyan, respectively). ] {\label{fig:ellipsoidsignals}\includegraphics[width=0.32\textwidth]{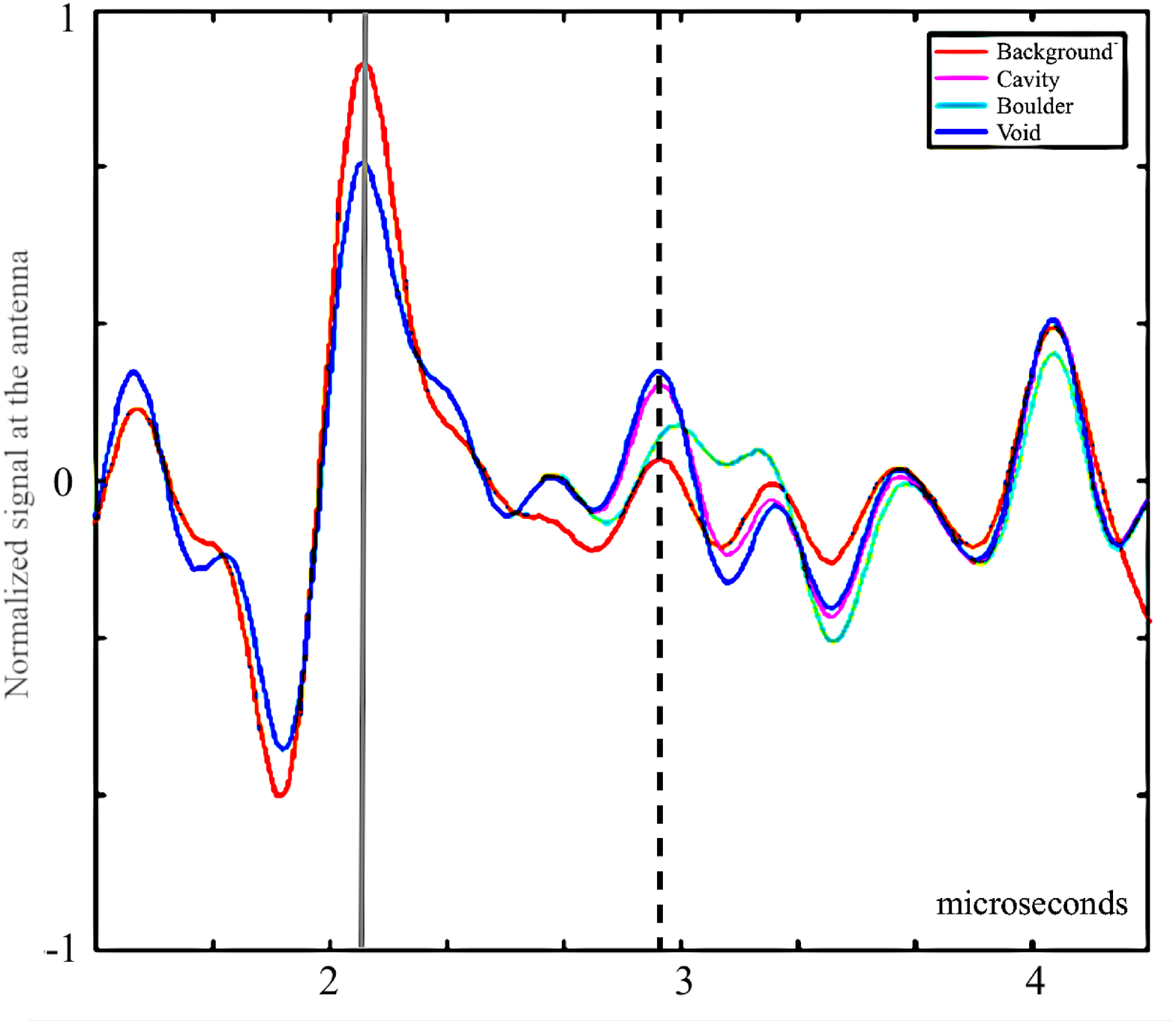}} \hskip1cm
\subfigure[Shallow and deep crack. In addition to the background signal ${\bf y}^{(bg)}$ (red), the actual signal ${\bf y}$ is visualized for the model (D) = Deep crack and (E) = Shallow crack (blue and magenta, respectively). ]{\label{fig:crackssignals}
\includegraphics[width=0.32\textwidth]{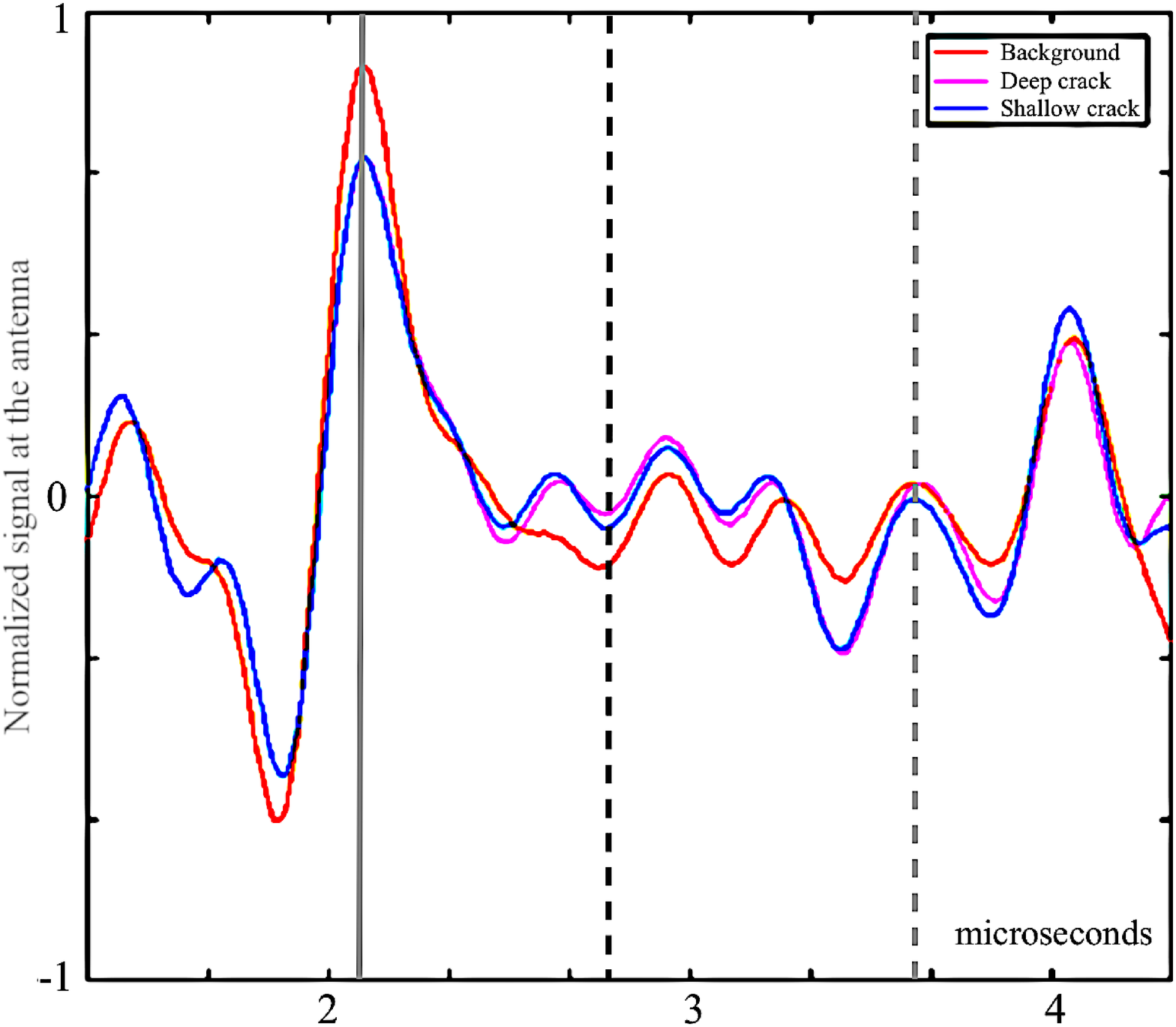}}
\end{center}
\caption{Time series of the normalized data at a single measurement point showing how the noiseless data of the models (A)--(E) differ in comparison to the simulated background signal (red). The effect of the surface layer on the data is marked by the solid vertical grey line. The dashed vertical black and grey lines on the right mark the effect of the shallow and the deep crack in the signal, respectively.}
\label{fig:signal}
\end{figure}

The full-wave computed radar tomography simulation was performed for  five different model structures (A)--(E) covering the following features: (A) a void space, (B) a highly porous inclusion, (C) a high-permittivity boulder, (D) a deep crack, and (E) a shallow crack inside the asteroid. The results of the simulations can be found in the Figures \ref{fig:signal}--\ref{fig:noise2} and the Table \ref{distinguishability_table}. The results suggest that deep interior permittivity anomalies can be detected inside a rubble-pile asteroid with limited-angle set of bistatic full-wave data. Nevertheless, detecting smooth structures was found to be difficult in comparison to how the void, boulder and crack details were distinguished. The model-specific results for (A)--(E) are described in the below sections. 

\subsection{Simulated measurement data}

Time series of the normalized data at a single measurement point is shown in the Figure \ref{fig:signal}. The shown time interval 1.3--4.3 $\mu$s shows signal propagation within the asteroid. Extending the duration beyond 4.3 $\mu$s was found to be unfeasible due to the noise effects caused by the inward-directed reflection peak from the outer surface facing away from the measurement position.  The first echo at time point 2 $\mu$s originates from the surface of the asteroid. In the background model (red line), the permittivity of the asteroid interior is constant $\varepsilon_r = 4$ across the whole asteroid interior and hence a more distinct echo is recorded in comparison to each the detailed model (blue line) which includes a surface layer with a lower permittivity value ($\varepsilon_r = 3$). Hence, also the amplitude of the reflected signal is lower in the latter case.

The figures \ref{fig:ellipsoidsignals} and \ref{fig:crackssignals} show how ellipsoids and cracks, respectively, can be distinguished in the raw signal at a single  measurement point. The differences in the curves are due to the differences in how the signal propagates through the asteroid. The non-linearity of the radar signal propagation is revealed by a mutual comparison between the amplitudes of the signals for models (A), (B), and (C). The geometrical models of these are identical, but the signal curves, however, differ also regarding the shape of the measured signal. The differences between the signals for the deep and shallow crack (D) and (E)  (Figure \ref{fig:crackssignals}) are minor but discernible. The shallow crack produces a more distinct echo which can be observed at the time point of approximately 2.7 $\mu$s. 

\subsection{Model (A): Void detection} \label{sec:bistaticvoid}

\begin{figure}[!ht] 
\subfigure[Model (A): Exact.]{\label{fig:voidmodel_3d}\includegraphics[width=0.225\textwidth]{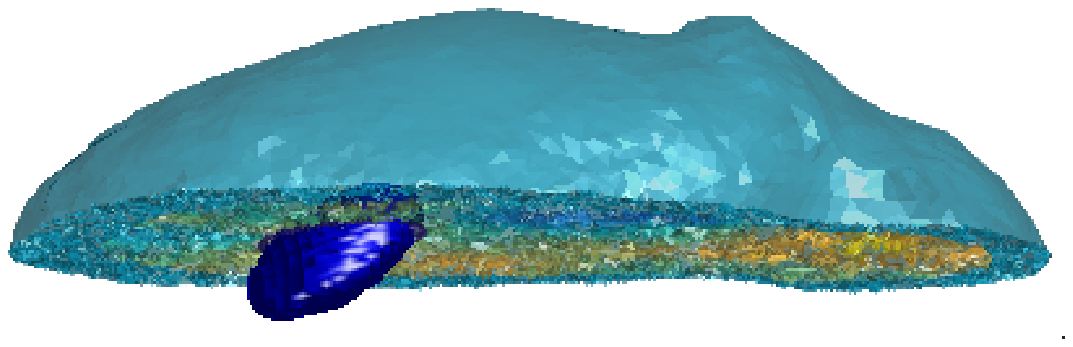}}
\subfigure[Model (A): 3D reconstruction of the detected void] {\label{fig:voidresults_3d}\includegraphics[width=0.225\textwidth]{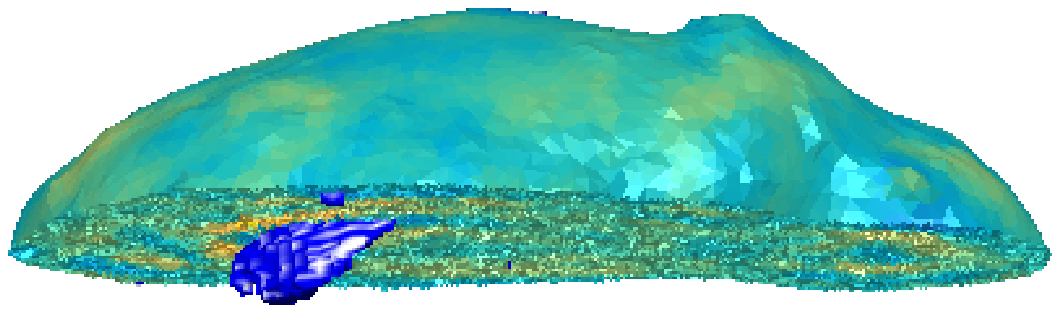}}

\subfigure[Model (A): Reconstruction after inversion showing also the surface layer in addition to the detection of the void.]{\label{fig:void_reconstruction} 
\includegraphics[width=0.45\textwidth]{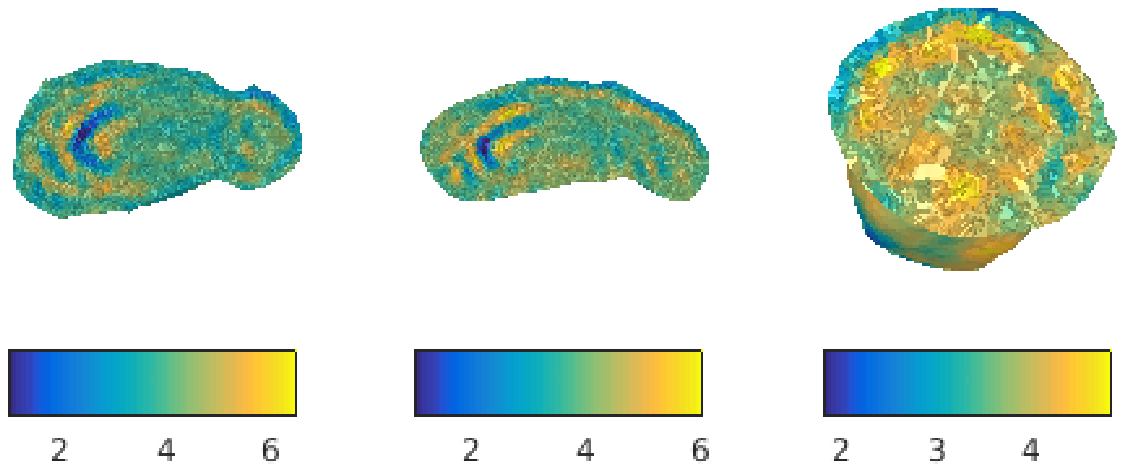}} 
\vskip1cm
\subfigure[Model (B): Exact.]{\label{fig:void2model_3d}\includegraphics[width=0.225\textwidth]{exact_model_3d_void2.eps}} 
\subfigure[Model (B): 3D reconstruction of the detected highly porous inclusion.] {\label{fig:void2results_3d}\includegraphics[width=0.225\textwidth]{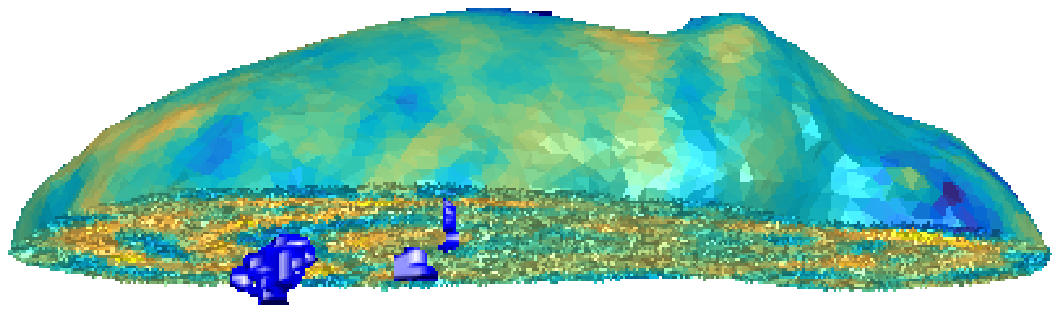}}
\hskip1cm
\subfigure[Model (C): Exact.]{\label{fig:void15model_3d}\includegraphics[width=0.225\textwidth]{exact_model_3d_void15.eps}}
 \subfigure[Model (C): 3D reconstruction of the detected high-permittivity boulder.] {\label{fig:void15results_3d}\includegraphics[width=0.225\textwidth]{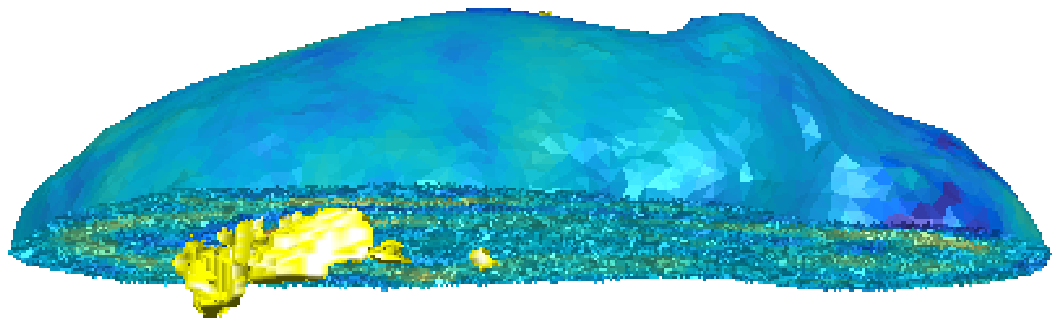}}

\subfigure[Model (B): Reconstruction after inversion showing also the surface layer. The highly porous cavity can be seen as enclosed by a higher permittivity layer (yellow).]{\label{fig:void2_reconstruction}
\includegraphics[width=0.45\textwidth]{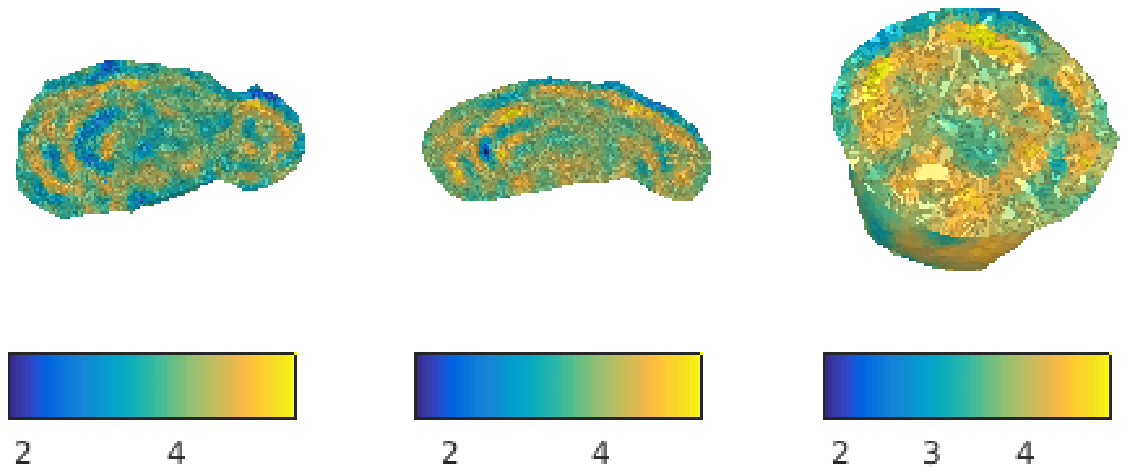}}
\hskip1cm 
 \subfigure[Model (C): Reconstruction after inversion showing also the surface layer. The high permittivity shape is clearly visible. ]{\label{fig:void15_reconstruction}
\includegraphics[width=0.45\textwidth]{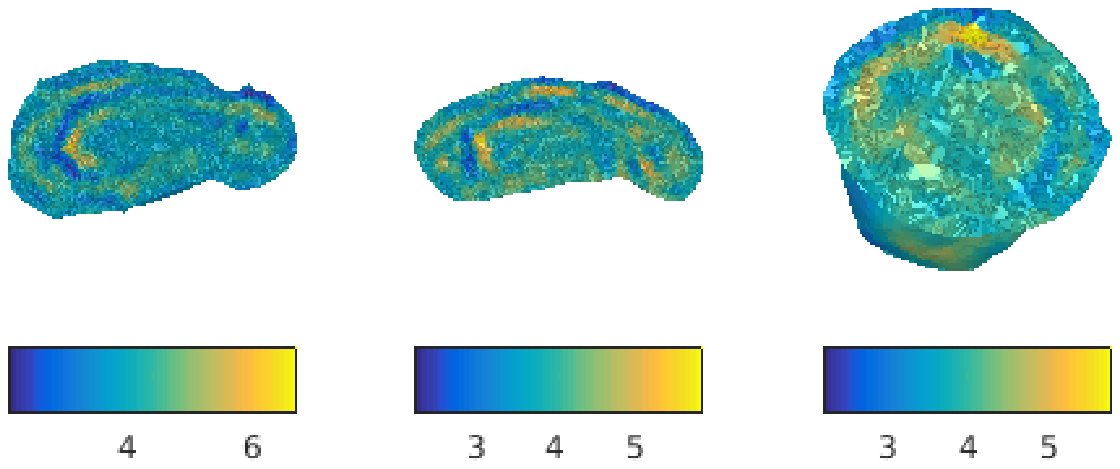}}
\caption{{\bf Top left:} The void of model (A) in (a) can be detected by full-wave computed radio tomography. 3D reconstruction of the detected void is shown in (b). The cut-view of the reconstruction in (c) shows a reconstruction which corresponds to a very realistic radargram in which the surface layer and the deep internal void can be detected. {\bf Bottom left:} For model (B),  a highly porous inclusion (d) is moderately visible in 3D (e) and cut-view reconstructions (h). {\bf Bottom right:} For model (C), the boulder with a high relative permittivity value in (f) ($\varepsilon_r = 15$) can be detected inside the asteroid in both the 3D (g) and cut-view (i) reconstructions.}
\label{fig:ellipsoid_results}
\end{figure}

A comparison between the reconstructed permittivity and  the exact 3D distribution for model (A) is shown in the Figures \ref{fig:voidmodel_3d} and \ref{fig:voidresults_3d}, respectively. It is evident that the (vacuum) void space of (A) is detectable. 
The quality of the radial (depth)  accuracy with respect to the asteroid's body was observed to be better than in the tangential one. This is obviously due to the sparse spatial distribution of the measurement points compared to the relatively dense time-resolution of the data which results in reliable localization capability. 

\begin{figure}[!ht] 
 \subfigure[Model (D): Exact deep crack.]{\label{fig:crackmodel_3d}\includegraphics[width=0.225\textwidth]{3d_deep_crack.eps}}
 \subfigure[Model (D): The detected crack in the deep interior.] {\label{fig:crackresults_3d} \includegraphics[width=0.225\textwidth]{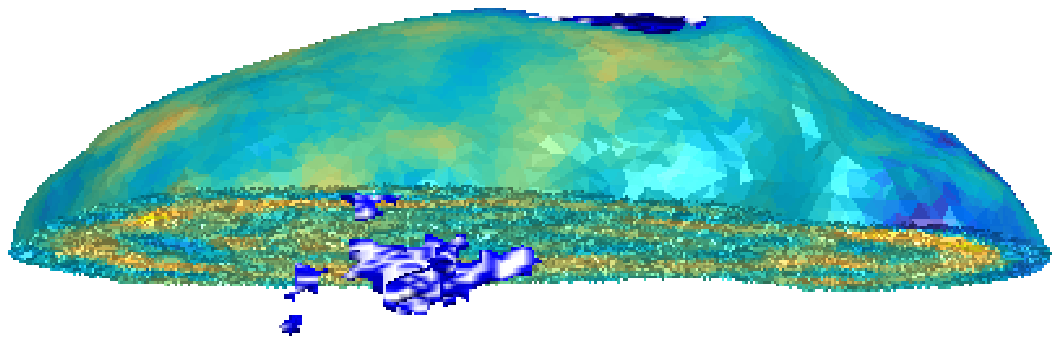}} \hskip1cm 
 \subfigure[Model (E) Exact shallow crack.] {\label{fig:crackxmodel_3d}\includegraphics[width=0.225\textwidth]{3d_crackx.eps}}
\subfigure[Model (E): The detected shallow crack.] {\label{fig:crackxresults_3d}\includegraphics[width=0.225\textwidth]{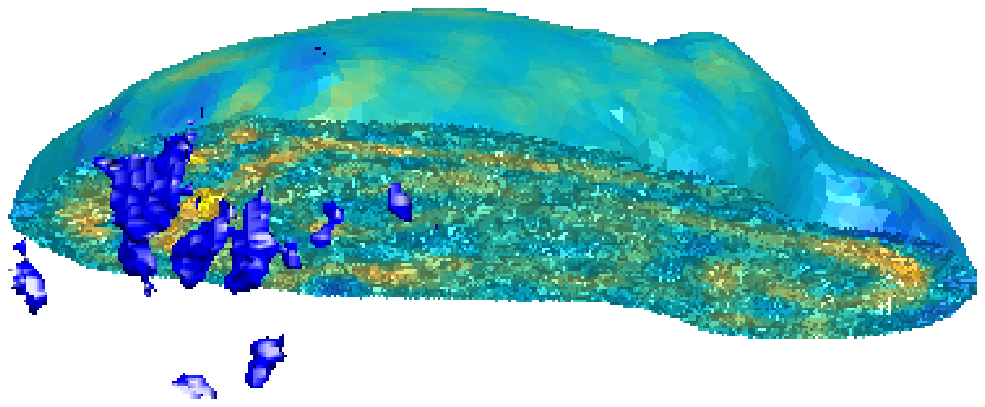}} \\ 
 \subfigure[Model (D): Deep crack model reconstruction after inversion showing also the surface layer in addition to the detection of the crack.]{\label{fig:crack_reconstruction}
\includegraphics[width=0.45\textwidth]{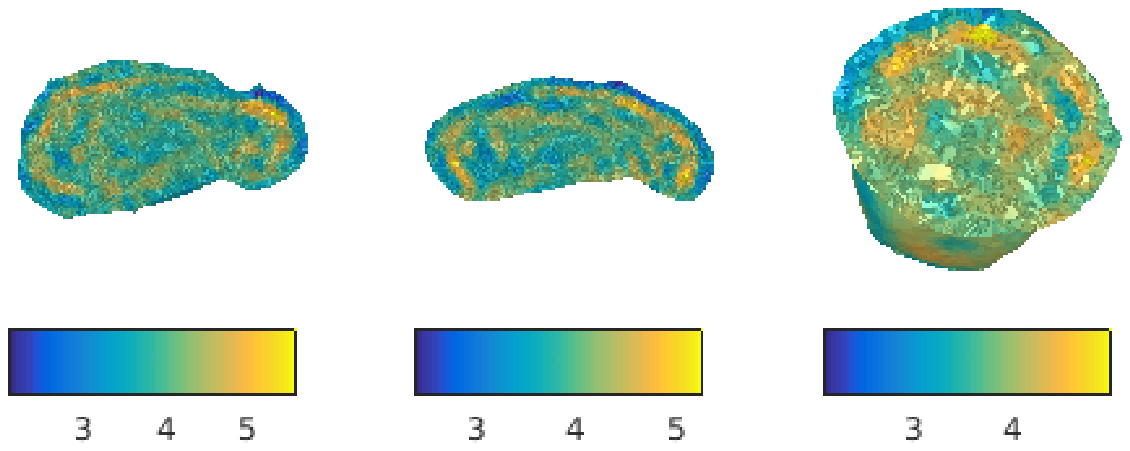}}  \hskip1cm 
\subfigure[Model (E): Shallow crack model reconstruction after inversion showing also the surface layer in addition to the detection of the crack.]{\label{fig:crackx_reconstruction}
\includegraphics[width=0.45\textwidth]{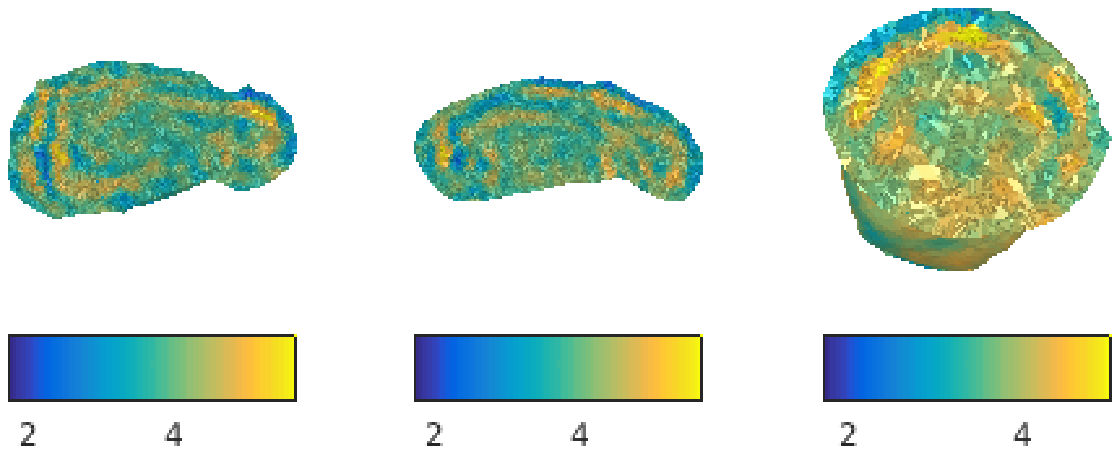}}
\caption{{\bf Left:} For model (D), the deep crack running across the asteroid in the vertical direction (a) is practically absent in the 3D (b) and cut-view (e) reconstruction. {\bf Right:} For model (E), the shallow crack with two branches and a complex shape (c) can be detected in the 3D (d) and cut-view (f) reconstructions.\label{fig:crackx_results}}
\label{fig:crack_results}
\end{figure}

\subsection{Model (B): Detection of a highly porous inclusion}

The task of detecting  highly porous materials ($\varepsilon_r = 2$), case (B), was also found to be feasible. However, the reconstruction quality obtained for (B) was weaker compared to that of (A)  (Figure \ref{fig:void2results_3d}). The exact shape of the inclusion found for (B) is not as obvious as in the case of (A), but the location of the anomaly is evident as shown by the Figures \ref{fig:void2results_3d} and \ref{fig:void2_reconstruction}. 

\subsection{Model (C): Boulder detection}

The Figures \ref{fig:void15results_3d} and \ref{fig:void15_reconstruction} show that a high-permittivity boulder can be detected. Similar to the void detection, also here the best detection accuracy was obtained in the transversal direction. The surface layer in the reconstruction (Figure \ref{fig:void15_reconstruction}) is less obvious than for the cases of (A) and (B). 

\subsection{Model (D): Deep crack detection}

 The reconstruction for model (D) shown on the left in the Figure \ref{fig:crack_results} suggests that the the deep crack is practically invisible for the radar. Obviously, its location in the deep interior,  and the limited-angle data, i.e., the absence of the measurement points around the z-axis, are potential factors limiting the detectability of the crack.

\subsection{Model (E): Shallow crack detection}

In contrast to the deep crack, a moderately clear echo was detected for the shallow crack in the case (E) as shown on the right in the Figure \ref{fig:crack_results}. Based on the reconstruction (Figure \ref{fig:crackxresults_3d}), the complex shape can also be discerned. However, the exact shape and, especially, the branches of the crack cannot  be accurately distinguished. 

\subsection{Goodness of reconstructions}

The qualitative assessment of the goodness of the reconstructions presented in the Figures \ref{fig:ellipsoid_results} and \ref{fig:crack_results} is shown in the Table \ref{distinguishability_table}.
The quantitative goodness of a reconstruction is measured by the mean square error (MSE) and the mean absolute error (MAE) computed separately for the anomaly detail and the surface layer, in addition to the global reconstruction containing the detail, the surface layer and the remaining asteroid interior of the asteroid model. The relative mean absolute error (MAE-R) is computed for the details only to compare the goodness of the detail reconstruction between the models (A)--(E).  

\textbf{
\begin{deluxetable*}{lllll}
\tablecaption{Distinguishability of model details \label{distinguishability_table}}
\tablecolumns{5}
\tablenum{2}
\tablewidth{0pt}
\tablehead{
\colhead{} &
\multicolumn{2}{c}{Surface layer}& \multicolumn{2}{c}{Interior}\\ \colhead{Model} &  \colhead{Data\tablenotemark{a}} & \colhead{Reconst.\tablenotemark{b}} & \colhead{Data\tablenotemark{a}} & \colhead{Reconst.\tablenotemark{b}} 
}
\startdata
(A) & Clear & Moderate & Clear & Clear \\
(B) &  Clear & Moderate & Clear & Moderate\\ 
(C) & Clear & Moderate & Moderate  & Moderate\\ 
(D) & Clear & Moderate & Weak & Weak \\
(E) & Clear & Moderate & Weak & Moderate \\
\enddata
\tablenotetext{a}{Based on the data curve in Figure \ref{fig:ellipsoidsignals}.}
\tablenotetext{b}{Based on the reconstructions in Figures \ref{fig:ellipsoid_results} and \ref{fig:crack_results}.}
\end{deluxetable*}
}

The computed mean and absolute error values (Table \ref{table:mse_table}) reflect the qualitative reconstructions. The errors are greatest in the detail areas and smallest in the surface layer. 
As is evident in the reconstruction images in the Figures \ref{fig:ellipsoid_results} and \ref{fig:crack_results}, the majority of the volumes in the ellipsoids and the cracks is not captured by the reconstruction and hence the permittivity values of these areas differ from the exact model very clearly. This is especially the case with the boulder model (C), in which the difference between the permittivity values in the exact model and the reconstruction varies greatly.

\begin{deluxetable*}{llccc}[b!]
\tablecaption{Mean square errors (MSE), mean absolute errors (MAE), and relative mean absolute errors (MAE-R)
\label{table:mse_table}}
\tablecolumns{5}
\tablenum{3}
\tablewidth{0pt}
\tablehead{
\colhead{Model} & \colhead{Error} & \colhead{Detail} & \colhead{Surface layer} & \colhead{Global} }
\startdata
(A) & MSE & 8.21 & 0.99 & 1.19 \\
 & MAE & 2.80 & 0.89 & 0.93 \\
 & MAE-R\tablenotemark{a} & 0.93 & \dots & \dots\\
(B) & MSE & 3.65 & 0.89 & 0.98 \\ 
  & MAE & 1.90 & 0.92 & 0.91 \\
  & MAE-R\tablenotemark{a} & 0.95 & \dots & \dots \\
(C) & MSE & 124.67 & 0.92 & 3.53\\ 
  & MAE & 11.12 & 0.87 & 1.08 \\
  & MAE-R\tablenotemark{a} & 1.01 & \dots & \dots \\
(D) & MSE & 7.10 & 0.93 & 1.05 \\
  & MAE & 2.49 & 0.89 & 0.90 \\
  & MAE-R\tablenotemark{a} & 0.83 & \dots & \dots \\
(E) & MSE & 6.86 & 0.90 & 1.04 \\
 & MAE & 2.38 & 0.87 & 0.89 \\
 & MAE-R\tablenotemark{a} & 0.79 & \dots & \dots
\enddata
\tablenotetext{a}{Normalized relative to the difference between the detail permittivity and the global mean of the Gaussian random field (4).}
\tablecomments{The category ''Detail'' refers to the ellipsoidal anomaly in the models (A)--(C) and to the deep and shallow crack in the models (D) and (E), respectively. The category ''Global'' refers to the complete asteroid model.}
\end{deluxetable*}

\subsection{Comparison of bistatic to monostatic measurement}

Model (A) was used to compare reconstructions obtained with  monostatic (single-satellite) and bistatic measurement data.  The results show (Figures \ref{fig:bistatic_result} and \ref{fig:monostatic_result}) that both satellite configurations enable the detection of the void. The bistatic measurement appears to show slightly more prominent shape of the anomaly, but this may not be significant in practical applications. Overall, bistatic measurement approach appears to provide  robustness to the reconstruction process. The same finding was previously reported and quantified in \citet{farfield} and can be accounted for by the fact that, in addition to the second receiver, the bistatic measurement set-up includes also the monostatic transmitter-receiver data collection. Hence, the bistatic measurement includes more information on the object.

\subsection{Effect of noise}
\begin{figure}[!ht] 
\mbox{} \hskip0.475\textwidth 
\subfigure[Bistatic mode]{\label{fig:bistatic_result}
\includegraphics[width=0.225\textwidth]{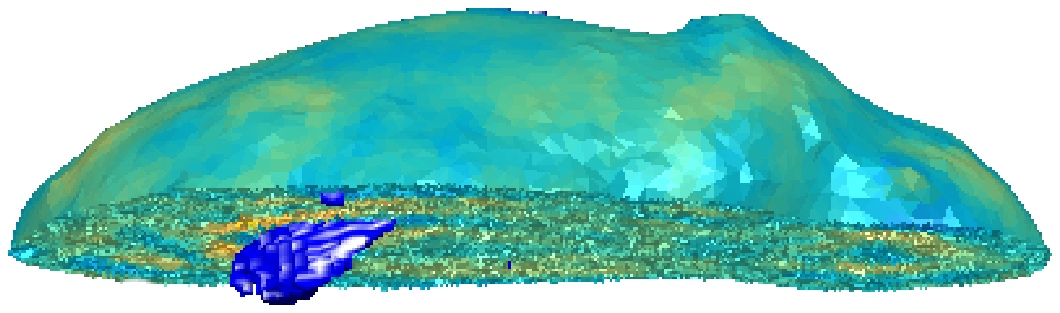}}
\subfigure[Monostatic mode]{\label{fig:monostatic_result}
\includegraphics[width=0.225\textwidth]{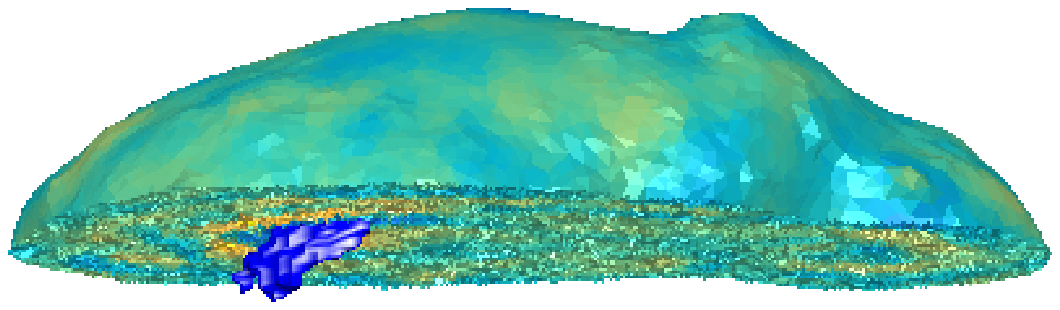}} 
\\
\vskip1cm
\subfigure[Noise -25 dB]{\label{fig:noise25}
\includegraphics[width=0.225\textwidth]{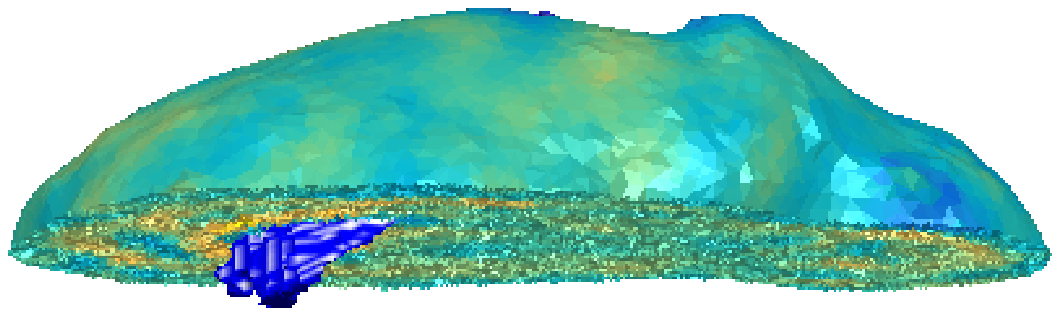}}
\subfigure[Noise -22 dB]{\label{fig:noise22}
\includegraphics[width=0.225\textwidth]{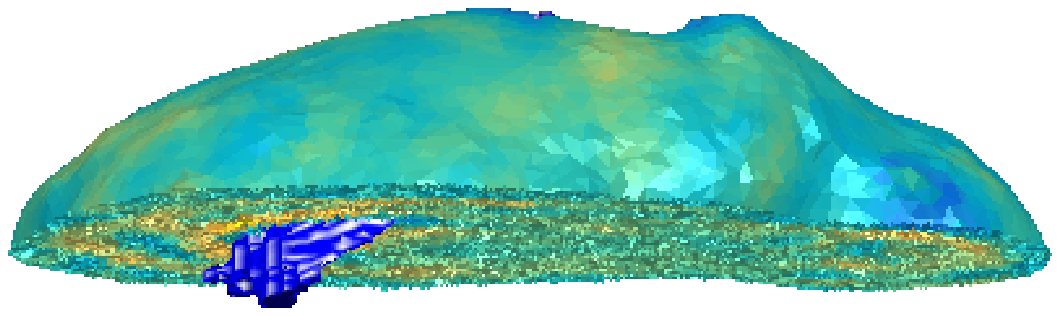}}
\subfigure[Noise -15 dB]{\label{fig:noise15}
\includegraphics[width=0.225\textwidth]{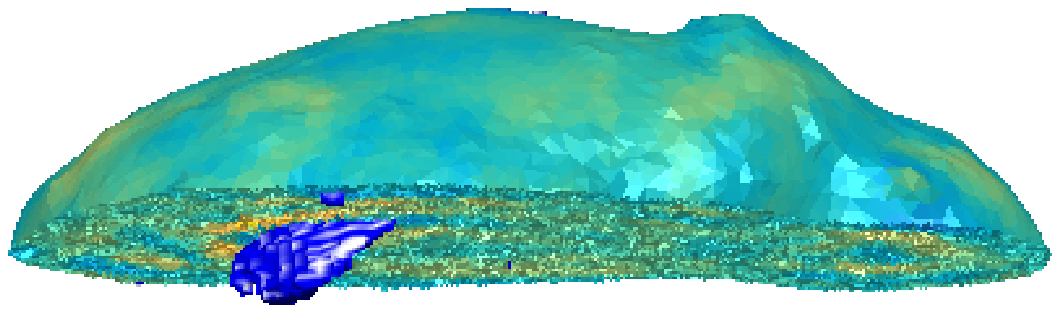}}
\subfigure[Noise -8 dB]{\label{fig:noise8}
\includegraphics[width=0.225\textwidth]{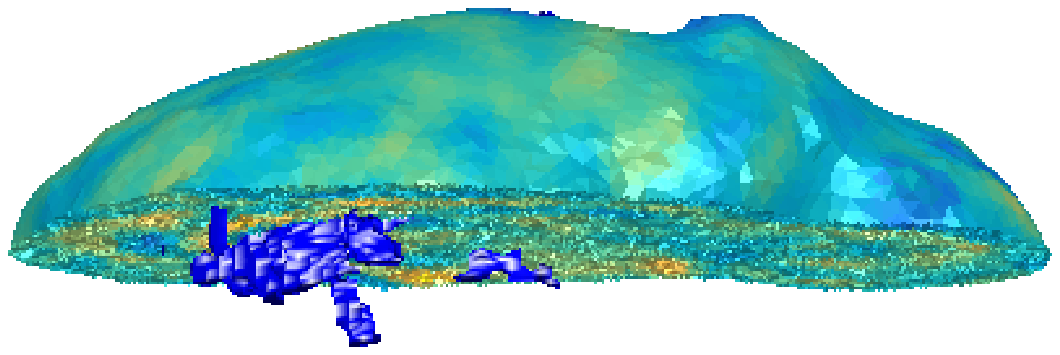}} \\
\subfigure[Noise -4 dB from two directions]{\label{fig:noise4}
\includegraphics[width=0.225\textwidth]{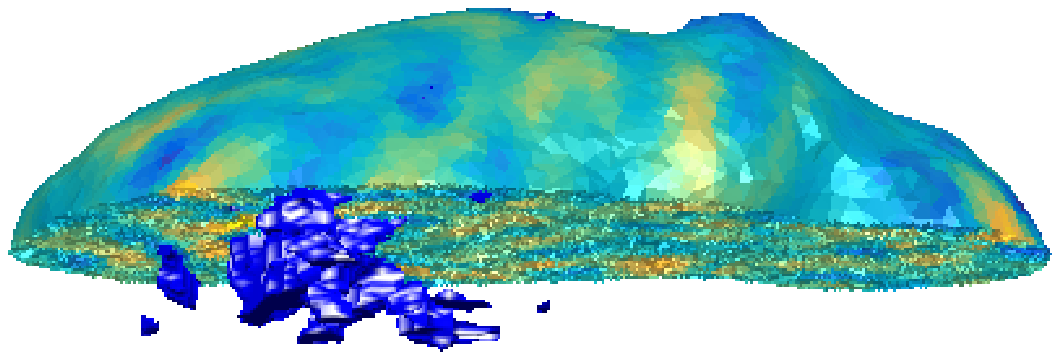}
\hskip0.05\textwidth
\includegraphics[width=0.125\textwidth]{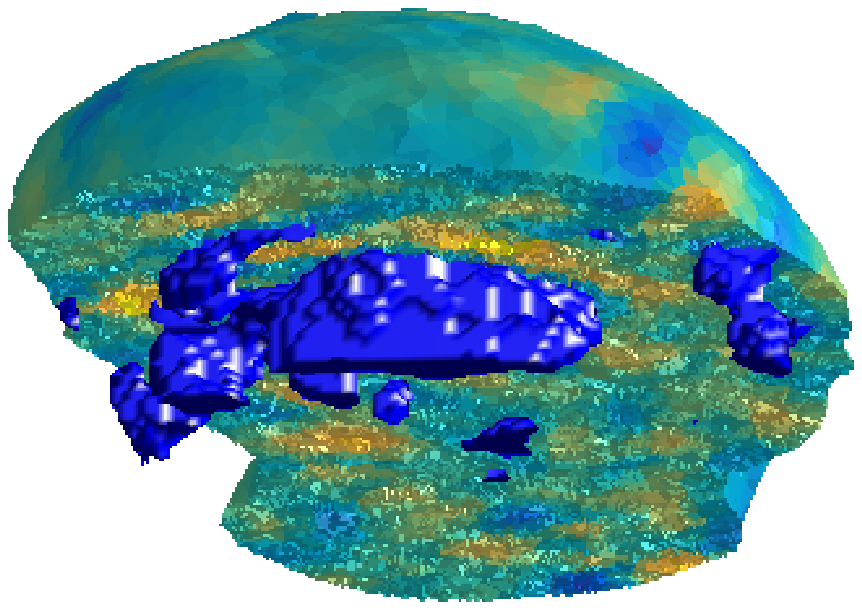}} \hskip0.05\textwidth
\subfigure[Noise -0 dB from two directions]{\label{fig:noise0}
\includegraphics[width=0.225\textwidth]{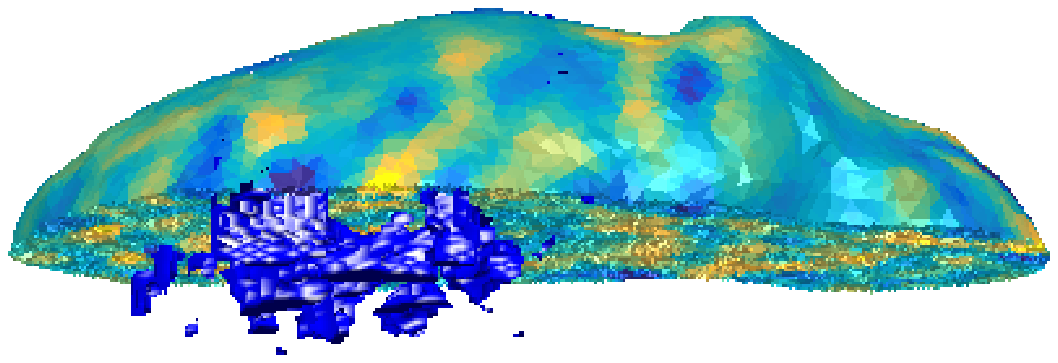}
\hskip0.05\textwidth
\includegraphics[width=0.125\textwidth]{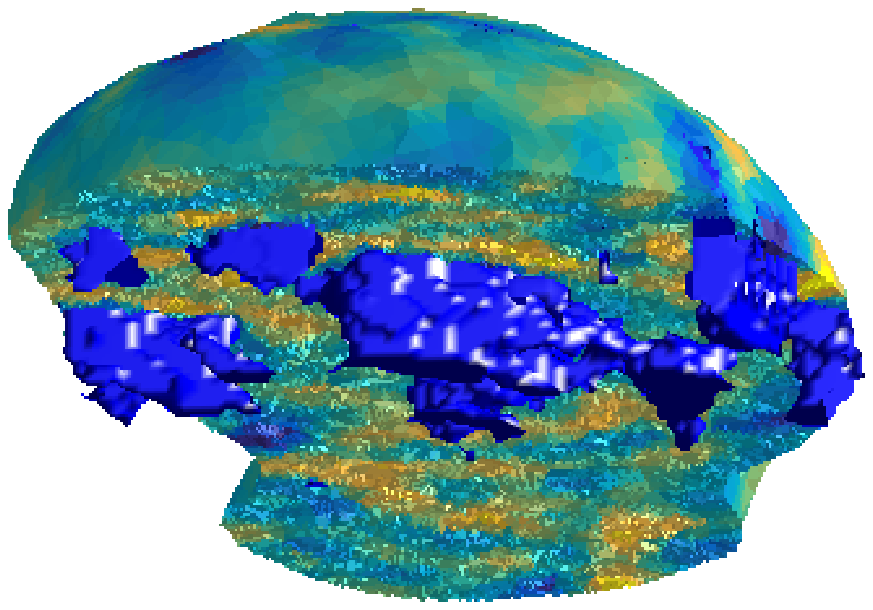}} \hskip0.05\textwidth
\caption{{\bf Top:} Comparison of bistatic (a, left) and monostatic (b, right) measurement configuration on the reconstruction of deep interior void. The bistatic set-up provides slightly better reconstruction. 
\label{fig:bivsmonostatic}
{\bf 2nd and 3rd row:} Detection of void (A) with different noise levels between -25 and 0 dB (c)--(h). The critical noise level losing the shape of the void is between -15 (e) and -8 dB (f)). Above -8 dB, the void can be moderately detected, but artifacts appear around the void area.}
\label{fig:noise1}
\label{fig:noise2}
\end{figure}

Finally, noise levels between -25 and 0 dB were investigated to determine the effect of noise on the reconstruction accuracy. Based on the results (Figures \ref{fig:noise25}-\ref{fig:noise0}), the artifacts appear at the -8 dB level after which they increase rapidly along with the noise. 

\section{Discussion}
\label{discussion}

The aim of this paper was to numerically validate the full-wave computed radar tomography (CRT) model developed by \citet{pursiainen2016} and \citet{farfield} for complex-structured asteroid models and especially to estimate the reconstruction capability of the CRT  for  rubble-pile asteroids. The surface model of the asteroid Itokawa was used because it is one of the few asteroids which the surface and bulk properties are known \citep{abe, fujiwara, okada, barnouinjha, nakamura, tsuchiyama}, and for which a high-resolution surface model is available \citep{itokawastlfile}. The structural models (A)--(E) of this study were generated by combining the data and knowledge  available on the physical properties of small solar system bodies and on how electromagnetic waves behave in such media \citep{kofman2012, kofman2015, herique2017}. 

To create a complex model for the deep interior part, roughly matching with the mass-concentration estimates obtained from the asteroid impact evolution studies \citep{Jutzi2017}, the dielectric properties of the asteroid were modelled by a Gaussian random field, which is a well-established approach to modelling random spatial structures in geostatistics and earth sciences \citep{geostatisticalsimulationbook,randomfieldsinearthsciencebook, geostatisticsbook}. Gaussian random fields are also used to generate random 3D porous structures \citep{Roberts2002} and modelling concrete \citep{concretemodelling} in civil engineering applications. Models (A)--(E) were obtained adding a surface layer together with deep interior details to a  Gaussian random field. Special interest was paid on modelling the inhomogeneous deep interior permittivity and void space accounting for bulk macroporosity of the body, which can be expected to be, for example, as high as 41 percent in the rubble-pile Itokawa \citep{fujiwara}.

The results suggest that permittivity anomaly details can be detected using CRT within a complex-structured asteroid, excluding the deep areas around the center of mass which are challenging due to signal attenuation and the limited-angle  measurement point distribution with an aperture around the z-axis. However, it also appears that the smooth Gaussian random field structure is difficult to reconstruct accurately. Compared to the  earlier work \citep{pursiainen2016, farfield},  the present inversion scheme is more realistic, showing that the practically achievable  reconstruction quality  will largely depend on the shape and the dielectric properties of the asteroid body. Moreover, it is obvious that the structures closer to the surface are detected more clearly than those lying deep inside the body. 

Reconstructing the interior details was shown to be feasible up to approximately -10 dB of noise. For a carrier frequency of 20 MHz, the galactic noise can be estimated to be around $5\times 10^{-20}$ W m$^{-2}$Hz$^{-1}$. At a distance of 1 AU, the dominating measurement noise is likely to originate from the Sun which radiates with the magnitude of $5\times 10^{-19}$ W m$^{-2}$Hz$^{-1}$ and $4\times 10^{-23}$ W m$^{-2}$Hz$^{-1}$ for its active and inactive (quiet) phase of sunspot activity, that is, for surface temperatures $10^6$ and $10^{10}$ K, respectively \citep{kraus1967, solarradio}. During the active phase there are radio emissions in time scales varying from seconds to hours. This noise can be diminished, by some amount, via orienting the antenna in an optimal way. When moving away from the Sun, the radiation, in general, decreases proportionally to the inverse square of the radius. This means that at 3 and 30 AU, the radiation levels due to the Sun will be -19 and -59 dB, respectively, with respect to those experienced at 1 AU. The latter one of these values is a rough estimate for the Kuiper belt.

The imaging resolution was found to be higher in the radial direction than in the tangential one. The details recovered also seem to be elongated in the tangential direction, exhibiting similar wave-front properties as conventional radargrams \citep{daniels2004}. The elongation can be interpreted to be due to the sparsity of the spatial measurement points. It is also present in other full-wave applications of CRT  \citep{gueting2015imaging,gueting2017high}, suggesting that enhancing the results significantly without changing the radar specifications might be difficult. Consequently, it seems that exact shapes cannot be accurately reconstructed, for example, structures oriented towards the center of the asteroid.  This is also reflected in the mean square errors and mean absolute errors computed for the reconstructions. However, based on the crack detection results it seems also reasonable to assume that  even complex  structures near the surface can be detected. The quality of reconstruction and the exact permittivity values may also be assumed to depend on the applied inversion technique and regularization parameters. In the present study, some of the details, most prominently the surface layer, are more obvious in the data (Figure \ref{fig:signal}) than in the final reconstructions (Figures \ref{fig:ellipsoid_results} and \ref{fig:crack_results}), suggesting that the inversion quality might be improved for example via depth weighting or noise-robust techniques such as multigrid, Markov chain Monte Carlo sampling or expectation maximization techniques  \citep{liu2008monte,tilley2017extending,multigrid}. 

Other future work on this topic will be to investigate methods to utilize complementary data in the inversion process. One way is to use multiple radar frequencies as suggested by \citet{herique2017}. For example, a frequency range of 25, 50 and 100 MHz can be considered to provide optimization of radar penetration depth. However, because the radar signal attenuation is directly proportional to signal frequency \citep{kofman2012}, the scattering and attenuation effects are stronger in the higher end of such a frequency range. In the present Itokawa model, extra signal attenuation in the deep part at the ellipsoidal void detail is 10 to 30 dB in the center frequency range of 50 to 100 MHz \citep{kofman2012}, respectively. Therefore, due to the lower penetration capability, it is possible that 50 and 100 MHz frequencies are suitable to exploring only the surface layer of asteroids and the deep interior details can only be recovered by low-frequency waves.  The analysis in this work involved only inversion of the modulated signal across the whole bandwidth. By combining the data from the modulated signal and the envelope, the accuracy of the method might be further improved, for example in improving crack detection. Moreover, gravimetry combined with radar sensing might help to improve the visibility of smooth structures or layers \citep{fregoso2009cross, park2014gravity}. Resolving the effect of surface scattering from anomalies such as large boulders, steep hills or crevasses on the quality of the inversion, and enhancing the noise model to distinguish different error sources, will also be important future work direction. 

\section{Conclusion}
\label{conclusion}

The results of this paper show that bistatic computed radar tomography can detect deep interior voids, cracks and high permittivity boulders inside a complex rubble pile  asteroid model when the total noise level in the data is around  -10 dB  with respect to the signal amplitude. The results suggest that permittivity anomaly details can be detected within a complex-structured asteroid by using CRT, excluding the deep areas around the center of mass which are challenging due to signal attenuation and the limited-angle  measurement point distribution with an aperture around the z-axis.

Furthermore, the bistatic measurement set-up improves the robustness of the inversion compared to the monostatic case. However, recovering the smooth Gaussian background distribution was found to be difficult with the present approach, suggesting that complementary techniques, such as gravimetry, might be needed to improve the reliability of the inference in practice. The inversion quality might be improved via method development and/or analyzing complementary data, such as a higher-frequency signal,  carrier wave envelope or gravity field measurements.  

\section{Acknowledgments}

L.-I.S., M.T.\ and S.P.\ were supported by the Academy of Finland Centre of Excellence in Inverse Modelling and Imaging. 

\appendix
\section{Computational framework and performance}\label{app:gpu}

For inverting the data, we used the computational full-wave radar tomography approach developed in \citet{pursiainen2016}, \citet{multigrid}, and \citet{farfield}. The computationally intensive part of the simulation was run in the GPU partition of Tampere Center for Scientific Computing (TCSC) Narvi cluster which consists of 8 GPU nodes with 20 CPU cores, 4 GPU each, totalling 32 NVIDIA Tesla P100 16 GB GPUs. The inversion reconstruction procedures were run on a Lenovo P910 workstation equipped with two Intel Xeon E5 2697A $2.6$ GHz 16-core processors and 128 GB RAM.

The creation of the system from node and tetrahedra data, including all parts of the asteroid geometry, orbiter points and interpolation of the finite element mesh required approximately 11 GB of memory space and \numprint{6700} seconds (approximately 1 hrs 52 minutes). Out of this, the actual system creation took 652 seconds (approximately 10 minutes) and creating the spatial interpolation matrix accounted for the rest of the computational time.

The minimum computation time of one transmitted signal was \numprint{11400} seconds (approximately 3 hrs 10 minutes). The system size was approximately $4.8$ GB. Because the forward simulation was parallelized into 16 processes, the total computation time of the entire simulation, 64 transmitter points, took a minimum of \numprint{45400} seconds (approximately 12 hrs 37 minutes). A typical expected computation time for the entire forward simulation was approximately $13.5$ hours.

The present study shows that full-wave (full-band\-width) data can be computed and inverted for a  rubble-pile asteroid with a realistic size and shape using a state-of-the-art computing cluster. GPU acceleration was observed to be necessary in the forward simulation stage in order to achieve feasible computation times and a speed-up by at least a factor of 10. In radar applications, similar experiences have recently been reported, for example in \citep{cordua2013appendix}. While the GPU-based forward simulation approach performs well regarding the speed, it is at the moment restricted with respect to  the system size. In our case, doubling the asteroid diameter would not have been possible due to the limited GPU RAM which in our cluster is currently 12 GB. Since GPUs are a rapidly progressing field of  technology, the available memory capacity may be expected to be significantly larger within a few years. Further development of the present computation framework may involve a pipeline for simulating realistic asteroid interiors based on a finite element mesh structural model.



\bibliographystyle{aasjournal}
\bibliography{refs.bib}







\end{document}